\documentclass[a4paper]{article}
\usepackage[english]{babel}
\usepackage[utf8x]{inputenc}
\usepackage{amsmath}
\usepackage{graphicx}
\usepackage[margin=3cm]{geometry}
\usepackage{setspace}
\usepackage[table,xcdraw]{xcolor}
\usepackage{caption}
\usepackage{subcaption}
\usepackage{booktabs}

\title{Statistical Methods for Selective Biomarker Testing}

\author{A. Adam Ding$^1$, Natalie DelRocco$^2$, and Samuel Wu$^2$
}

\date{\today}

\begin{document}
\doublespacing
\maketitle
%%%%% Preamble to the document ends here %%%%%%
% next we include the different sections.
% Each section is included in the document

\footnotetext[1]{Department of Mathematics, Northeastern University, Boston, MA, USA}

\footnotetext[2]{Department of Biostatistics, University of Florida, Gainesville, FL, USA\vspace{\baselineskip}\\
\textbf{Corresponding Author:}\\
Aidong Adam Ding, Department of Mathematics, Northeastern University, Boston, MA, USA\\
Email: a.ding@northeastern.edu
}

\begin{abstract}

Biomarker is a critically important tool in modern clinical diagnosis, prognosis, and classification/prediction. However, there are fiscal and analytical barriers to biomarker research. \textit{Selective Genotyping} is an approach to increasing study power and efficiency where individuals with the most extreme phenotype (response) are chosen for genotyping (exposure) in order to maximize the information in the sample. In this article, we describe an analogous procedure in the biomarker testing landscape where both response and biomarker (exposure) are continuous. We propose an intuitive reverse-regression least squares estimator for the parameters relating biomarker value to response. Monte Carlo simulations show that this method is unbiased and efficient relative to estimates from random sampling when the joint normal distribution assumption is met. We illustrate application of proposed methods on data from a chronic pain clinical trial.\vspace{\baselineskip}

\end{abstract}

\section{Introduction}
\label{s:intro}

A biological marker, or biomarker, is an objective measurement which indicates a biologic process or response \cite{Wagner2002OverviewDevelopment}. This umbrella definition captures a range of measurements that may be representative of a disease course, from simple indicators like the blood pressure to complex laboratory tests \cite{Strimbu2010WhatBiomarkers}. In this paper, we focus specifically on physical markers which are detectable via serum assays. In the past 20 years, the explosive use of biomarkers in medical research has coined the “biomarker revolution” \cite{Schisterman2012TheRevolution}. Clinically relevant biomarkers can provide information on both disease mechanisms and subsequent outcomes \cite{Atkinson2001BiomarkersFramework}. In practice, a biomarker can serve as both a risk indicator and a surrogate for disease status. As such, the biomarker is a critically important tool in modern clinical diagnosis, prognosis, and classification/prediction.

Despite the clinical utility and popularity of biomarkers and continual advancements in collection technology, there remain fiscal and analytical barriers to biomarker research.  The cost of conducting biomarker assays for a sufficiently powered study is a major limitation. For example, a luminex assay with 39 sample slots and the capacity to detect up to 20 biomarkers can cost between \$300 and \$600 before fees for consultation, lab materials, and labor \cite{ufAssayPrice2020}. This means that collecting biomarker samples for each person in a study of 800 participants could cost more than \$12,600 when serum samples for each individual are collected at a single time point. Resource limitations have thus inspired the development of cost-effective experimental designs and corresponding statistical methodology \cite{Schisterman2012TheRevolution, Albert2012NovelData}.

In genetics literature, one such approach concentrates sampling to the most informative observation units \cite{Darvasi1992SelectiveLocus}. Selective genotyping \cite{Lander2012MappingMaps} traditionally entails sampling individuals for genotyping based on extreme phenotypic values where genotype (exposure) is discrete (presence/absence; aa/Aa/AA) and phenotype (response) is continuous, most often assuming an ANOVA-style model \cite{Muranty1997SelectiveLocus}. From a statistical perspective, this is distinct from, but perhaps inspired by, the information perspective \cite{Sen2005QuantitativePerspective} for which the optimal design in regression analysis (i.e. sample) is the one that minimizes the variance of coefficient estimates \cite{Holt1980RegressionSurveys}.

Under selective genotyping, the response no longer follows a normal distribution and missing data in the middle of the phenotypic distribution must be accounted for \cite{Rabier2014OnGenotyping}. Appropriate maximum likelihood methods have been developed for inference under such designs for QTL studies \cite{Darvasi1992SelectiveLocus, Muranty1997SelectiveLocus}, which have been shown to increase statistical power relative to random sampling using a fraction of the original sample \cite{Carey1991LinkageSamples, Darvasi1992SelectiveLocus, VanGestel2000PowerTraits}. However, the developed methods are highly specific to the field of genetics, modeling the exposure as discrete and accounting for elements such as backcross in models. Additionally, in genotyping studies it is advantageous to take a multistage approach wherein promising genetic markers are identified early out of a pool of candidates to meet study constraints \cite{Satagopan2002Two-stageStudies, Satagopan2004Two-stageConstraints}.

In this paper we study a selective biomarker-testing scheme where, similar to selective genotyping, individuals with extreme response values are selected for biomarker-testing. In contrast to typing the discrete genotypes in the selective genotyping, here the individuals' continuous biomarkers are measured.
Selective genotyping and our selective biomarker-testing are both special cases of Outcome Dependent Sampling (ODS) \cite{Sen2005QuantitativePerspective}. ODS (also known as response-dependent sampling) is a form of biased sampling, which has been studied in the context of likelihood-based estimation  \cite{Lawless1999SemiparametricRegression}. Zhou et al.~\cite{Zhou2002AOutcome} formally defined an ODS design, for a linearly related continuous exposure and outcome where all response values are observed and covariate values are observed, as (1) a simple random sample (SRS) from the full available cohort and (2) additional SRS’s from regions of the response that are of particular interest.  They proposed a semiparametric empirical likelihood estimation method for this ODS design which was shown to increase efficiency relative to simple random sampling \cite{Zhou2002AOutcome}. Weaver and Zhou~\cite{Weaver2005AnSampling} extended this design to incorporate all available information from the full sample including those with unknown exposure values taking an estimated likelihood approach. Zhou’s original estimator has been expanded to accommodate different functional forms \cite{Tan2016EstimationOutcome}, auxiliary covariate information \cite{Wang2006ACovariates, Wang2009Outcome-Inference, Zhou2011StatisticalOutcome}, multi-stage designs \cite{Song2009AOutcome, Zhou2011StatisticalOutcome, Xu2012MixedOutcome}, mixed effects models \cite{Xu2012MixedOutcome}, and more recently survival \cite{Yu2021AcceleratedSampling} and longitudinal endpoints \cite{Schildcrout2012Outcome-dependentVariable, Albert2012NovelData, Zelnick2018Likelihood-basedData}.

Here we propose statistical analysis methods to analyze selective biomarker-testing data utilizing regression estimations available from standard statistical software.
Notice that our selective biomarker-testing scheme can be considered as Extreme Outcome Dependent Sampling (EODS) since we only sample individuals with extreme response values without the SRS from the full cohort (thus no biomarker-testing is conducted for any individual in the mid range of response values). The benefit, relative to the cost, of the incorporation of the primary SRS in traditional ODS designs is unknown. Selective genotyping literature suggests that there is no information to be gained by genotyping individuals outside the tails of the response variable distribution  \cite{Darvasi1992SelectiveLocus}. Also, while the ODS estimators in above literature are efficient, unbiased, and flexible under their statistical assumptions, the complexity of the semiparametric likelihood-based approach is a key barrier to their widespread use in practice. The least squares approach to estimation in regression analysis is standard, and therefore most accessible to researchers conducting biomarker-testing studies. However, least squares has been shown to be most susceptible to the bias induced by extreme sampling based designs \cite{Holt1980RegressionSurveys}. To solve this issue, we propose a reverse-regression least squares estimation method for EODS designs with jointly normal distributed biomarker and response.

The organization of this article is as follows: In Section 2 we describe the BISP2 clinical trial dataset which motivated this research. In Section 3 we describe a general methodology for biomarker studies, including biomarker selection, the effect estimation, power/sample size considerations, and model checking methods. In Section 4 we present numerical results of our estimator. We first study the finite sample properties of our estimator via simulation, then apply our estimator to the BISP2 Biomarker Study.

\section{Motivating Example}
\label{s:data}

Biopsychosocial Influence on Shoulder Pain Phase I (BISP) was a single-center, pre-clinical “proof of concept” study of 190 adults to identify genetic and psychological characteristics related to chronic musculoskeletal pain \cite{Borsa2018GeneticInjury}. Musculoskeletal pain is the general pain affecting the muscles, ligaments, tendons, or bones with chronic indicating that the pain is long-lasting or consistently recurring. In general, the musculoskeletal pain is a large contributor to the \$635 billion yearly healthcare cost of chronic pain in the United States \cite{Borsa2018GeneticInjury}. Though this makes chronic pain a high priority research area, there are limited accepted treatment models due to the complexity of disease etiology. Treatment components must be personalized on the basis of genetic, psychological, environmental, and social risk factors, which all contribute to the individual variation in how people experience chronic pain.

Specifically, BISP targeted chronic musculoskeletal pain affecting the shoulder region by comparing predictors of pain level among healthy individuals pre- and post-induced shoulder pain. The target population was healthy adults. Participants were followed over the course of five days. Baseline covariates and DNA samples were collected on the first day of the study, before inducing shoulder pain. An exercise-based protocol was then used to create controlled shoulder pain through inflammation and muscular fatigue \cite{Borsa2018GeneticInjury}. Four follow-up visits were conducted every 24 hours post-injury induction. Shoulder impairment, genetic testing, and other covariates related to pain level defined a priori based on clinical expertise were measured at baseline and each follow-up visit. BISP identified multiple prognostic factors which were associated with increased shoulder pain, including promising genes which showed evidence of predicting shoulder impairment.

BISP Phase II (BISP2) was a single-center, randomized follow-up trial to BISP which aimed to test whether interventions personalized on the basis of genetic and psychological characteristics are effective for induced shoulder pain \cite{George2022BiopsychosocialInjury}. The two-factor factorial design randomized 261 individuals to either propranolol or placebo and psychological education or general education. Propranolol is a drug chosen to target Catechol-O-methyl-transferace (COMT), which metabolizes adrenal hormones and is associated with pain sensitivity. The psychological education was designed to target pain rumination, which is magnification of pain by focusing on the pain with a pessimistic attitude. Pre-randomization, shoulder injury was induced using the same protocol.

BISP2 participants provided daily report on pain intensity and disability over the 5-day onsite observation period. Pain level was measured using the Brief Pain Inventory (BPI) \cite{Keller2004ValidityPain}, which is an 11-point scale ranging from 0 (no pain) to 10 (worst pain intensity imaginable). Participants rated the intensity of current pain and pain intensity at its worst, best, and average over the past 24 hours. Clinically relevant covariates and saliva samples for genetic testing were again taken at each follow up visit.

The BISP2 investigators targeted 14 genetic markers associated with pain in the study’s exploratory biomarker analysis. These biomarkers were chosen a priori based on clinical knowledge of propensity to release pro-inflammatory cytokines. All genetic samples were saliva-based. The primary question of interest for the exploratory analysis was: which biomarkers are associated with recorded level of shoulder pain at 48-hours post-randomization? As the response, pain, is measured via the BPI to represent a continuum pain, linear regression with pain level as the response is a typical analysis approach. However, the investigators were limited by the budget and time constraints of the BISP2 clinical trial. It was infeasible to biomarker-test every individual in the study. Therefore, the investigators began by biomarker-testing 31 individuals in the high and low tails of worst pain experienced at 48-hours post shoulder pain induction. Determining an appropriate analysis method for data from this extreme sampling motivates the methodology in the remainder of this paper.

\section{Methodology}
\label{s:methods}

In a clinical trial, we are interested in the relationship between a response variable $Y$ and a biomarker variable $X$. If both variables are measured on all subjects enrolled in the trial, we have a full data set consisting of $(X_1,Y_1)$, ..., $(X_{n_F}, Y_{n_F})$. Here $n_F$ denote the size of the full data set. For the EODS biomarker analysis, we observe the response variable for the full data set, $Y_1$, ..., $Y_{n_F}$, but only observe biomarker variable $X$ on a subset corresponding to extreme values of $Y$.
Say, we select $\gamma$ proportion of the full data set and measure the biomarker $X$ on the selected subset of size $n_S = [\gamma n_F]$, where $[d]$ denotes the closest integer to $d$.
Without loss of generality, we can denote the first $n_S$ subjects as the selected ones. That is, in the selective biomarker analysis, we observe $(X_1,Y_1)$, ..., $(X_{n_S}, Y_{n_S})$ and $Y_{n_S+1}, ..., Y_{n_F}$, while $Y_1$, ..., $Y_{n_S}$ corresponds to the top $\gamma/2$ proportion and the bottom $\gamma/2$ proportion of $Y_1$, ..., $Y_{n_F}$.

The aim of the statistical regression analysis is to study the dependence of response variable $Y$ on the biomarker $X$:
\begin{equation}\label{eq:LR.expect}
E(Y|X) = \alpha_Y + \beta_Y X,
\end{equation}
where $E(Y|X)$ denotes the conditional expectation of $Y$ given $X$. When $X$ and $Y$ jointly follows a bivariate normal distribution, the full data set satisfy the standard regression model assumption:
\begin{equation}\label{eq:LR.model}
Y_i = \alpha_Y + \beta_Y X_i + \varepsilon_{Y,i}, \qquad \varepsilon_{Y,i} \sim N(0, \sigma_{\varepsilon_Y}^2),
\end{equation}
for $i=1,..., n_F$.

\begin{figure}
    \caption{Illustration of challenge posed to linear regression under extreme sampling.}
    \label{fig:extremeSampVisual}
     \centering
     \begin{subfigure}{0.75\textwidth}
         \centering
         \caption{Truncation of the conditional distribution of $Y$ given $X$ under extreme response selection with the linear model approach.}
         \includegraphics[width=\textwidth]{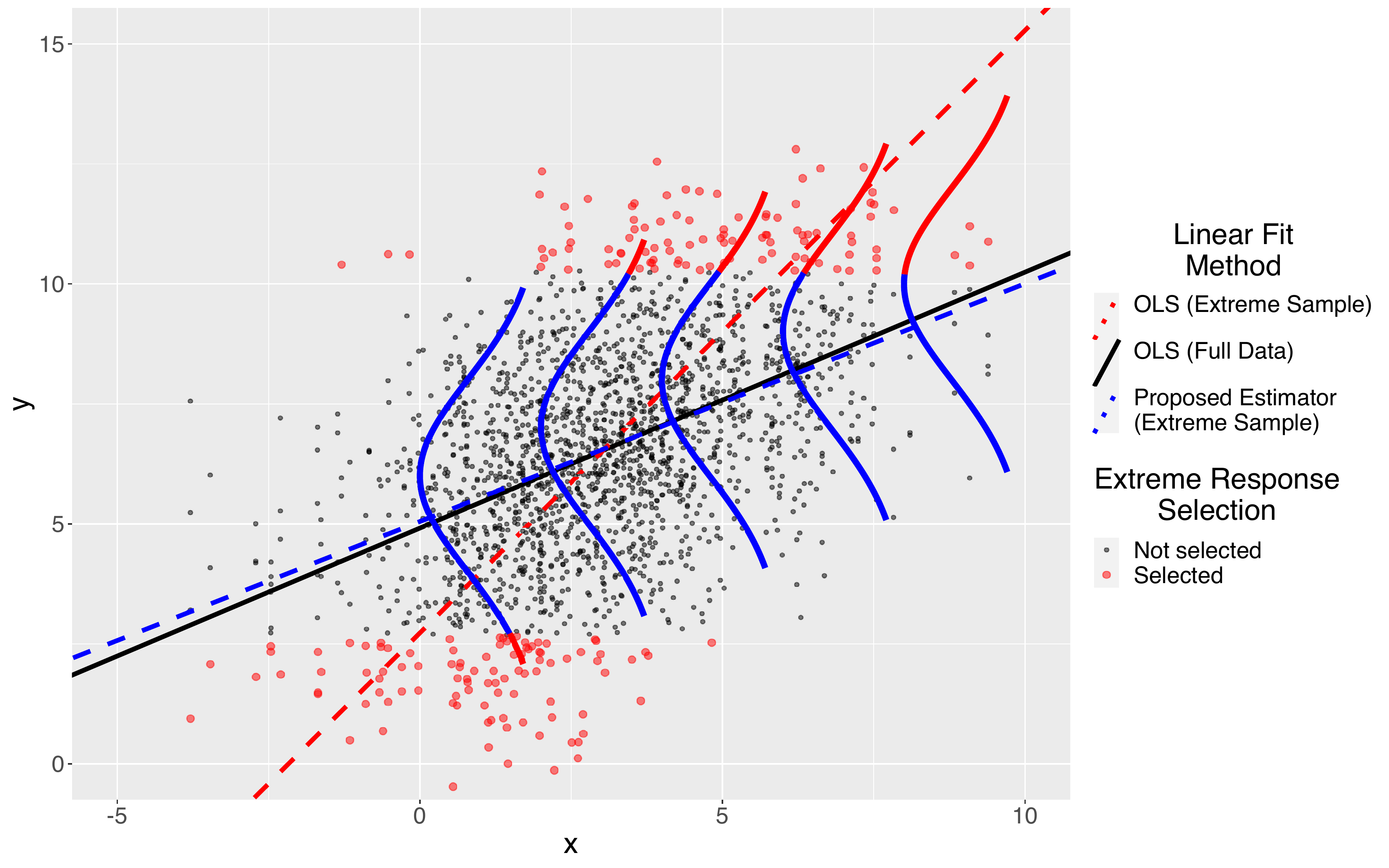}
         \label{fig:ygivenx}
     \end{subfigure}
     \par\bigskip
     \begin{subfigure}{0.75\textwidth}
        \centering
        \caption{Preservation of the conditional distribution of $X$ given $Y$ under extreme response selection.}
        \includegraphics[width=\textwidth]{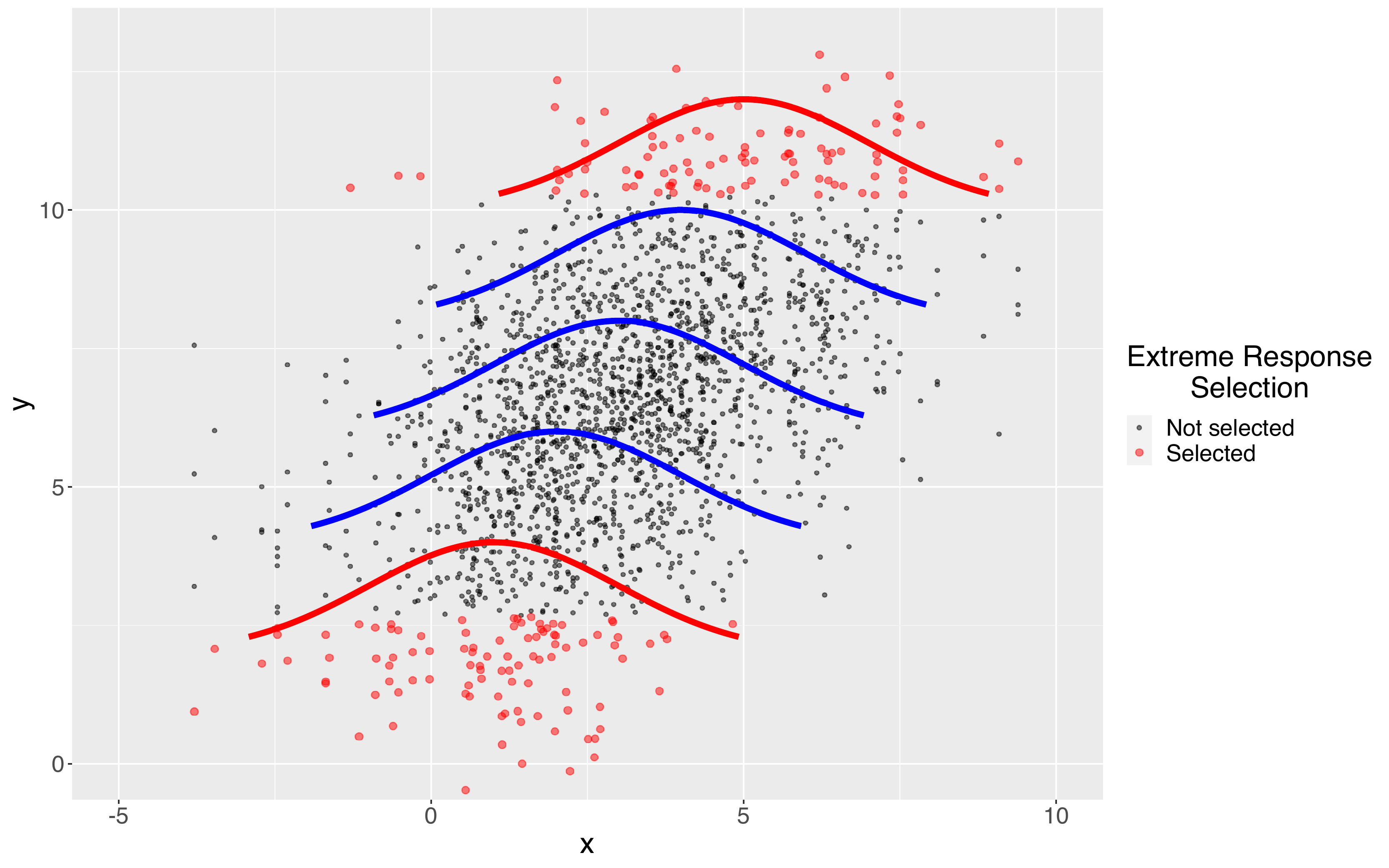}
        \label{fig:xgiveny}
     \end{subfigure}
\end{figure}

However, for the EODS biomarker analysis, naively conducting linear regression analysis directly on the subset $(X_1,Y_1)$, ..., $(X_{n_S}, Y_{n_S})$ does not work. The reason is that, the regression model assumption~\eqref{eq:LR.model} does not hold on the selectively sampled subset because the response follows a truncated distribution due to the selection of extreme values for $Y$ as shown by the red part of the curves in Figure~\ref{fig:ygivenx}. The ordinary least square (OLS) fit on the selectively sampled subset will lead to a highly unbiased slope estimation as shown by the dotted red line.
Therefore, specific analysis methods are needed for the selective biomarker analysis. In this paper, we propose a parametric analysis method which utilizes standard regression formulas on reverse-regression: regressing $X$ on $Y$ for the selectively sampled subset. The main insight for the proposal is that, when $X$ and $Y$ jointly follows a bivariate normal distribution, the standard regression model assumption holds for the reverse-regression on both full data set and the selectively sampled subset:
\begin{equation}\label{eq:LR.model.XonY}
X_i =  \alpha_X + \beta_X Y_i + \varepsilon_{X,i}, \qquad \varepsilon_{X,i} \sim N(0, \sigma_{\varepsilon_X}^2).
\end{equation}

As shown in Figure~\ref{fig:xgiveny}, conditional on selected $Y$ values, $X$ is still normally distributed. Armed with this insight, we can conduct the reverse-regression $E(X|Y) = \alpha_X + \beta_X Y$ using standard statistical software on the selectively sampled subset. Then we convert the reverse-regression fit results into regression inferences, with a little help of the additional response variable observations $Y_{n_S+1}, ..., Y_{n_F}$. The conversion formulas are described in detail in the rest of this section, with the mathematical derivations provided in the Appendix.

\subsection{Hypothesis Test}
\label{s:test}

One main focus of the biomarker analysis is to test whether a biomarker $X$ affects the response variable $Y$. Statistically, this is usually done by the linear regression hypothesis testing
\begin{equation}\label{eq:H0.YonX}
H_0: \beta_Y = 0 \qquad \mbox{versus} \qquad H_A: \beta_Y \ne 0.
\end{equation}

Mathematically, the null hypothesis $H_0: \beta_Y = 0$ is equivalent to the reverse-regression null hypothesis  $H_0: \beta_X = 0$. Thus we can simply carry out the reverse-regression hypothesis test on the selectively sampled subset for
\begin{equation}\label{eq:H0.XonY}
H_0: \beta_X = 0 \qquad \mbox{versus} \qquad H_A: \beta_X \ne 0.
\end{equation}
The p-value for the test of \eqref{eq:H0.XonY} is also valid for testing \eqref{eq:H0.YonX}.

\subsection{Biomarker Selection}
\label{s:selection}

When multiple biomarkers $X^{(1)},...,X^{(M)}$ are under study, testing is done for the $M$ biomarkers on the selectively sampled subset: $(X^{(1)}_1,...,X^{(M)}_1,Y_1)$, ..., $(X^{(1)}_{n_S},...,X^{(M)}_{n_S}, Y_{n_S})$. The usual variable selection methods can be applied on $X^{(1)},...,X^{(M)}$ using the p-values from the reverse-regression. That is, for each $j=1,...,M$, a p-value $p^{(j)}$ is yielded from testing
\begin{equation}\label{eq:H0.XonYj}
H_0: \beta_{X^{(j)}} = 0 \qquad \mbox{versus} \qquad H_A: \beta_{X^{(j)}} \ne 0
\end{equation}
on the selectively sampled subset. The biomarkers $X^{(j)}$'s can then be sorted in an order of increasing p-values $p^{(j)}$ with smaller p-value indicating stronger biomarker effect on the response. The strongest biomarkers can be selected for further study. To adjust for the multiple testing, common variable selection rules such as the  Benjamini–Hochberg method~\cite{BH1995} can be applied on these p-values.

\subsection{Effect Estimation}
\label{s:estimation}

The effect of the biomarker $X$ on response variable $Y$ is measured by the slope $\beta_Y$ in the regression function. Naively fitting the regression equation~\eqref{eq:LR.expect} on the selectively sampled subset would result in an overestimation of the biomarker effect. As illustrated in Figure~\ref{fig:ygivenx}, the conditional (truncated) distribution of $Y$ changes with the given $X$ values, leading to OLS estimate (shown as the dotted red line) with much bigger slope magnitude than the true $\beta_Y$ value when $\beta_Y > 0$. For consistent estimation of $\beta_Y$, we use instead the parameter estimates $\hat \beta_X$ and $\hat \sigma^2_{\varepsilon_X}$ from fitting the reverse-regression \eqref{eq:LR.model.XonY} on the selectively sampled subset. These estimates can be obtained from conducting the reverse-regression using any standard software. Furthermore, from the full response variable observations of $Y_1$, ..., $Y_{n_F}$, we have the sample mean $\tilde \mu_Y = \frac{1}{n_F} \sum_{i=1}^{n_F}Y_i$ and the sample variance $\tilde \sigma^2_Y= \frac{1}{n_F-1} \sum_{i=1}^{n_F} (Y_i - \tilde \mu_Y )^2$.

Using equation~\eqref{eq:BetaY_BetaX} in the Appendix, we have an point estimation for $\beta_Y$ as
\begin{equation}\label{eq:hatBetaY}
\hat \beta_Y  = \frac{1}{\frac{\hat \sigma^2_{\varepsilon_X}}{\tilde \sigma^2_Y} + \hat \beta^2_X } \hat \beta_X.
\end{equation}

Then the standard error for $\hat \beta_Y$ can be calculated as in equation~\eqref{eq:SE_BetaY},
$$
s.e.\{\hat \beta_Y\} = \sqrt{\frac{(\frac{\hat \sigma^2_{\varepsilon_X}}{\tilde \sigma^2_Y}-\hat \beta^2_X)^2 (s.e.\{\hat \beta_X\})^2 + (\frac{2 \hat \beta^2_X \hat \sigma^4_{\varepsilon_X}}{\tilde \sigma^4_Y})(\frac{1}{n_s-2} + \frac{1}{n_F-1})} {(\frac{\hat \sigma^2_{\varepsilon_X}}{\tilde \sigma^2_Y} + \hat \beta^2_X )^4}},
$$
whose detailed derivation is provided in the Appendix. Therefore, the $(1-\alpha)$ confidence interval for $\beta_X$ can be calculated as
$$
\hat \beta_Y \pm t_{\alpha/2, df=n_{S}-2} s.e.\{\hat \beta_Y\}.
$$
Here $t_{\alpha/2, df=n_{S}-2}$ is the $\alpha/2$-upper quantile for the t-distribution with $n_{S}-2$ degrees of freedom.

Similarly, using \eqref{eq:alphaY_BetaX} in the Appendix, $\alpha_Y$ is estimated by
\begin{equation}\label{eq:hatAlphaY}
\hat \alpha_Y  = \frac{\hat \sigma^2_{\varepsilon_X} \tilde \mu_Y - \hat \alpha_X  \hat \beta_X{\tilde \sigma^2_Y}}{{\hat \sigma^2_{\varepsilon_X}} + \hat \beta^2_X {\tilde \sigma^2_Y} }.
\end{equation}

%Description of the unbiased estimator $\hat{\beta}_x$ of $\beta_x$ and its corresponding standard error.

\subsection{Power/Sample Size Calculation}
\label{s:sample size}

As derived in the Appendix, equation~\eqref{eq:powerSel} gives the power formula of the hypothesis test in section~\ref{s:test}:
\begin{equation}\label{eq:powerS}
power = P(W > FQ_{\alpha, df_1, df_2}),
\end{equation}
where $W$ denotes a random variable following a non-central F-distribution with degrees of freedoms of $df_1 = 1$ and $df_2 = \gamma n_F-2$ and noncentral parameter $n_F f^2 2 \int_{z_{\gamma/2}}^{\infty} x^2 \phi(x) dx$, and $FQ_{\alpha, df_1, df_2}$ denotes the $\alpha$-upper quantile of a central F-distribution with degrees of freedoms of $df_1$ and $df_2$.
Here $z_{\alpha}$ denotes the $\alpha$-upper quantile of a standard normal distribution $N(0,1)$ whose density is denoted as $\phi(x)$. And $f$ is the Cohen's effect size defined as
\begin{equation}\label{eq:f^2}
f^2 = \frac{R^2}{1-R^2} = \frac{\rho^2}{1-\rho^2},
\end{equation}
where $R^2$ is the proportion of variation in the data explained by the regression equation.

Based on this, we can choose the proportion $\gamma$. We illustrate the sample size calculation with a simple example here. Assume that we have a full data set of sample size $n_F=200$, and we wish to detect a Cohen's effect size $f=0.3$ with $90\%$ power at the significance level $\alpha=0.05$. When we select $10\%$ individuals in EODS for biomarker-testing (i.e., $n_S=200*0.1=20$, selecting $10$ persons with largest $Y$ values and $10$ persons with smallest $Y$ values), \eqref{eq:powerS} gives $power=0.9150$. When we select $9\%$ individuals in EODS for biomarker-testing (i.e., selecting $9$ persons with largest $Y$ values and $9$ persons with smallest $Y$ values), \eqref{eq:powerS} gives $power=0.8985$. Thus $20$ individuals will be biomarker-tested to achieve the design goal of $90\%$ power.

In contrast, the standard t-test in simple linear regression for testing \eqref{eq:H0.YonX} has $power=0.8983$ when $n=118$ and $power=0.9007$ when $n=119$ as given by equation~\eqref{eq:powerFull}. Thus a clinical trial without the EODS will require $119$ individuals with both $X$ and $Y$ measurements, needing an almost $6$-fold increase in cost of biomarker-testing for $X$ than the EODS method.

%$6$-fold expensive biomarker-testing for $X$ than the EODS method.

\subsection{Model Checking}
\label{s:model.check}

Our methodology is based on the assumption that $X$ and $Y$ are jointly normally distributed. Mathematically, that is equivalent to the following two assumptions holding simultaneously: (A) the response variable $Y$ is normally distributed and (B) conditional on $Y$, the biomarker variable $X$ is normally distributed. We now consider how to check these two assumptions on observed data.

Assumption (A) can be checked with standard methods such as the normal probability plot on the fully observed $Y_1$, ..., $Y_{n_F}$. For assumption (B), we do not observe $X$ for the whole range of $Y$. But we can still check whether (B) holds for the selected $Y$ values by applying standard model checking methods on the reverse-regression, e.g., the normal probability plot of the reverse-regression residuals.

\section{Numerical Studies}
\label{s:results}

\subsection{Simulation Studies}
\label{s:simulations}

In this section, we use simulations to check the finite sample performance of our proposed Outcome-Dependent Extreme biomarker-testing (ODEB) estimator. All simulations were conducted using R Statistical Software Version 3.6.0 \cite{rCore2021}.

\subsubsection{Simulation Comparison with Ordinary Least Square}

We compare our ODEB estimation with the ordinary least square (OLS) estimation in various parameter settings here.
Data was simulated according to the probability model described in Equation~\eqref{eq:LR.model}. Specifically, $Y_i$ is generated such that
\begin{equation}\label{eq:Sim.model}
Y_i = 5 + \beta_Y X_i + \varepsilon_{Y,i}, \qquad \varepsilon_{Y,i} \sim N(0, 5).
\end{equation}
Individual data sets consisting of $(X_1,Y_1)$, ..., $(X_{n_F}, Y_{n_F})$ were randomly generated for each of $B=20,000$ Monte Carlo simulations. We studied combinations of the following sets of parameter values: $n \in \{100, 200, 400, 800\}$, $\beta_{Y} \in \{0,0.2,0.26,0.4,0.8,1\}$, $\gamma \in \{0.1,0.2,0.4\}$. For each iteration, both the extreme-outcome-dependent sampling and random sampling were conducted. $\beta_Y$ was then estimated for each sampling method using both OLS ($\hat{\beta}_{Y,OLS}$) and ODEB ($\hat{\beta}_{Y,ODEB}$) estimation.

From the theory in previous section~\ref{s:methods}, our ODEB estimation should work well for both the extreme sampling and random sampling  since the reverse-regression model assumption~\eqref{eq:LR.model.XonY} hold in both cases. OLS would work well for the random sampling, but would provide inconsistent estimation under the extreme sampling scheme. The simulation results confirmed this.

\begin{figure}
 \centering
  \caption{Bias of $\hat{\beta}_Y$ under 20\% of the total study sampled for various $\beta_Y$'s.}
	\includegraphics[width=\textwidth]{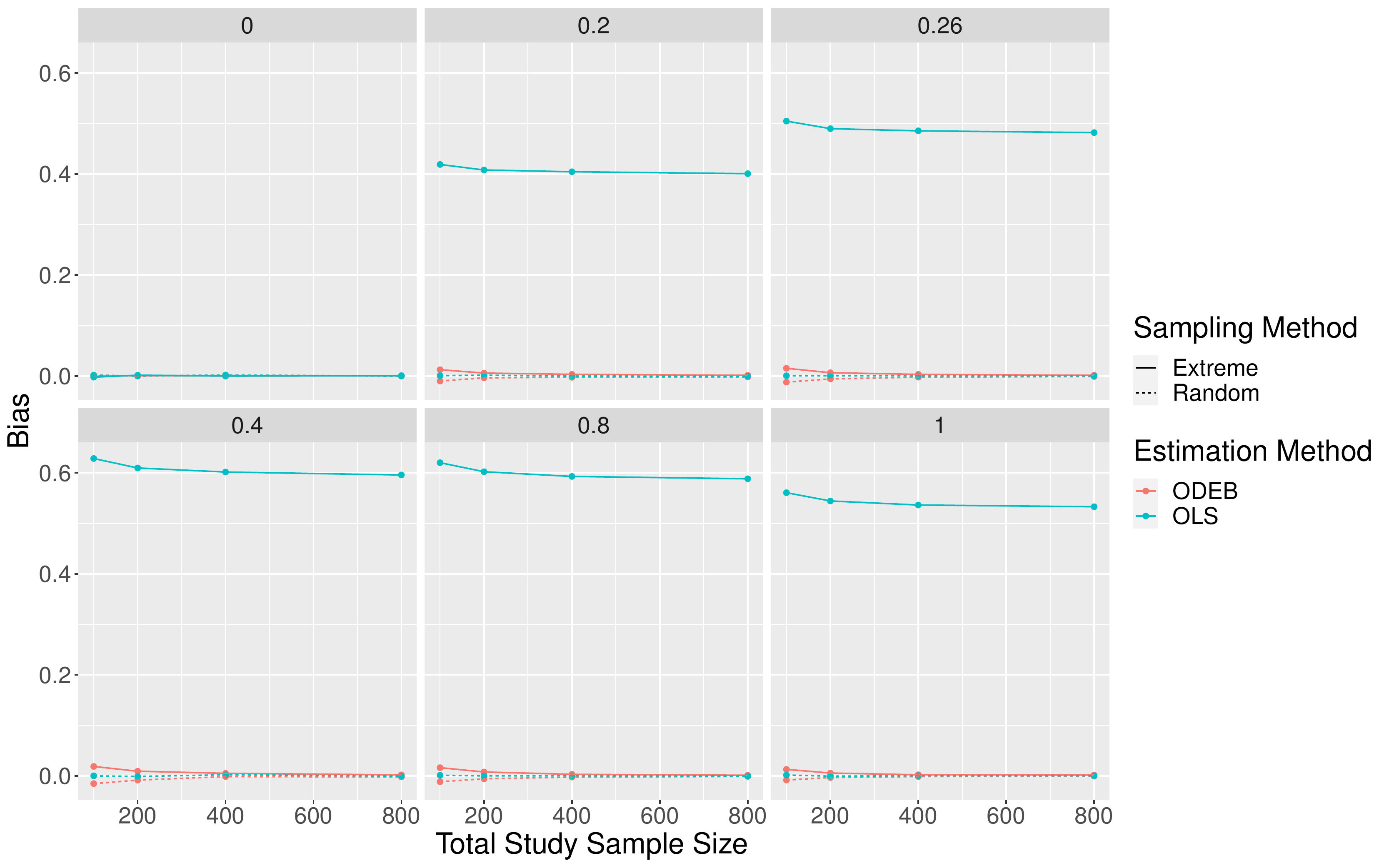}
 \label{fig:Norm_Bias_20per}
\end{figure}

Figure \ref{fig:Norm_Bias_20per} displays the bias of $\hat{\beta}_Y$ when an extreme sample of 10\% is taken from each tail of the response distribution (thus resulting in a total of 20\% extreme sampling). When there is truly no effect of the biomarker $X$, both OLS and ODEB estimation are unbiased regardless of whether the sample is extreme or random. When there is a biomarker effect present, $\hat{\beta}_{Y,ODEB}$ under extreme or random sampling and $\hat{\beta}_{Y,OLS}$ under random sampling remain mostly unbiased, with only a small appreciable bias when $n=100$ which dissipates as the sample size reaches $n=400$. However, $\hat{\beta}_{Y,OLS}$ under extreme sampling is severely biased in the positive direction. The bias initially increases, then tapers, with the size of $\beta_Y$.
For saving space, we omits the detailed numbers for simulations from other sampling proportions (10\% and 40\%) but summarize their bias patterns.
The bias of $\hat{\beta}_{Y,OLS}$ under extreme sampling is exacerbated when $\gamma$ decreases to 10\% (i.e., 5\% per tail), with estimation bias exceeding twice the magnitude of $\beta_Y$. The magnitude of the $\hat{\beta}_{Y,OLS}$ bias decreases slightly when $\gamma = 0.4$, reflecting contribution from more moderate observations, but the overall bias magnitude is still unacceptably large.

%%%%%%%%%%%%%%%%%%%% ODEB VS. OLS BIAS %%%%%%%%%%%%%%%%%%%%%%%%

\begin{figure}
 \centering
  \caption{Root mean square error (RMSE) of $\hat{\beta}_Y$ under 20\% of the total study sampled for various $\beta_Y$'s.}
	\includegraphics[width=\textwidth]{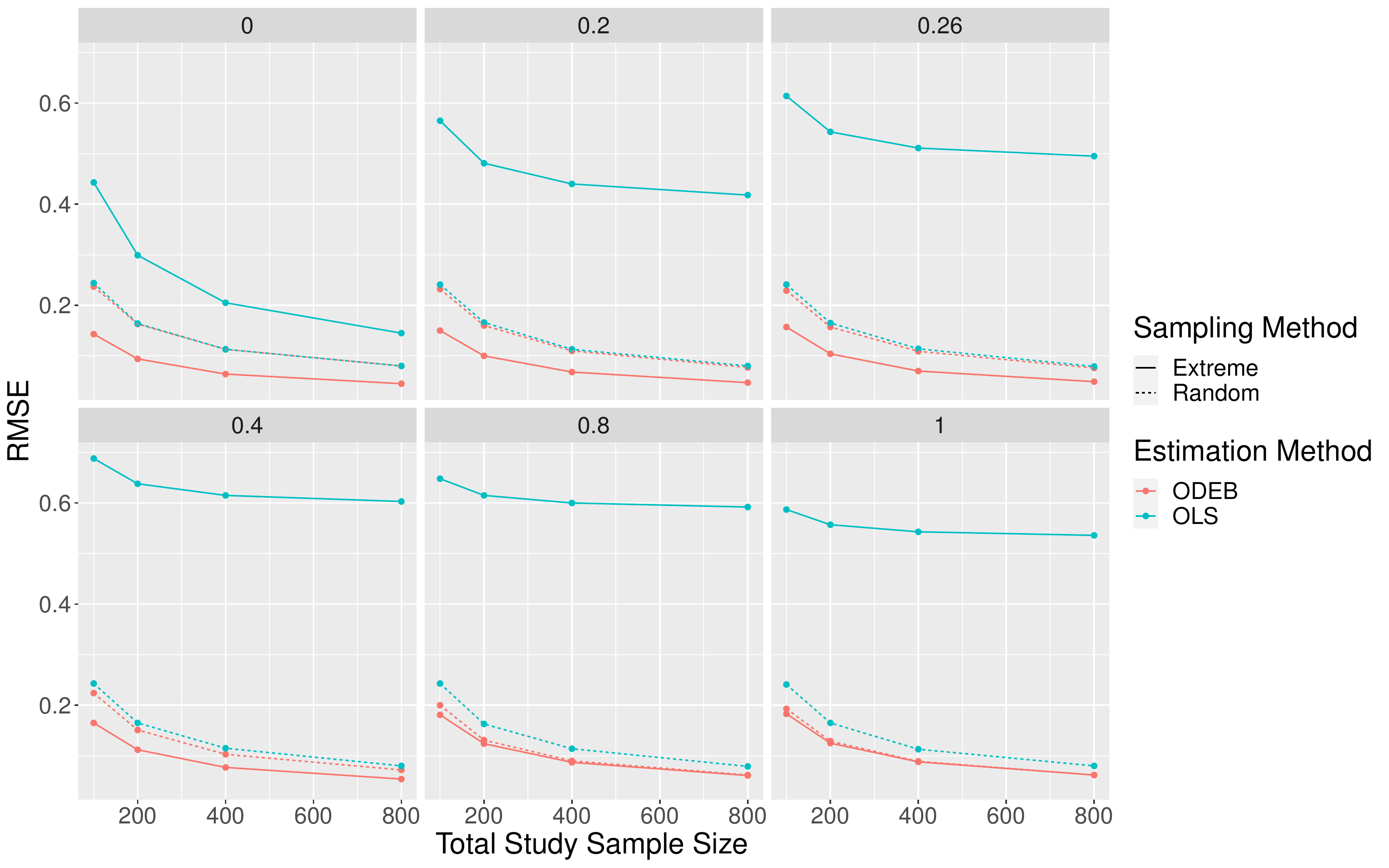}
 \label{fig:Norm_MSE_20per}
\end{figure}

The root mean square error (RMSE) displays similar trends, as shown in Figure \ref{fig:Norm_MSE_20per}. Under extreme sampling, $\hat{\beta}_{Y,ODEB}$ has the smallest RMSE of the four considered combinations across both study sample sizes and true $\beta_Y$'s. This reflects the efficient use of information by $\hat{\beta}_{Y,ODEB}$, which incorporates information from  observations with medial response values only through  the response variance estimation in equation~\eqref{eq:hatBetaY}. In contrast, $\hat{\beta}_{Y,OLS}$ has the largest RMSE across study sample sizes and $\beta_Y$'s when applied to an extreme sample. Of note, when biomarker-testing is conducted via random sample, $\hat{\beta}_{Y,OLS}$ and $\hat{\beta}_{Y,ODEB}$ often perform similarly in terms of RMSE, but $\hat{\beta}_{Y,ODEB}$ beats $\hat{\beta}_{Y,OLS}$ as $\beta_Y$ becomes bigger.

%%%%%%%%%%%%%%%%%%%% ODEB VS. OLS RMSE %%%%%%%%%%%%%%%%%%%%%%%%

\begin{figure}
 \centering
  \caption{Power to detect a nonzero effect of biomarker level on pain for various $\beta_Y$'s.}
	\includegraphics[width=\textwidth]{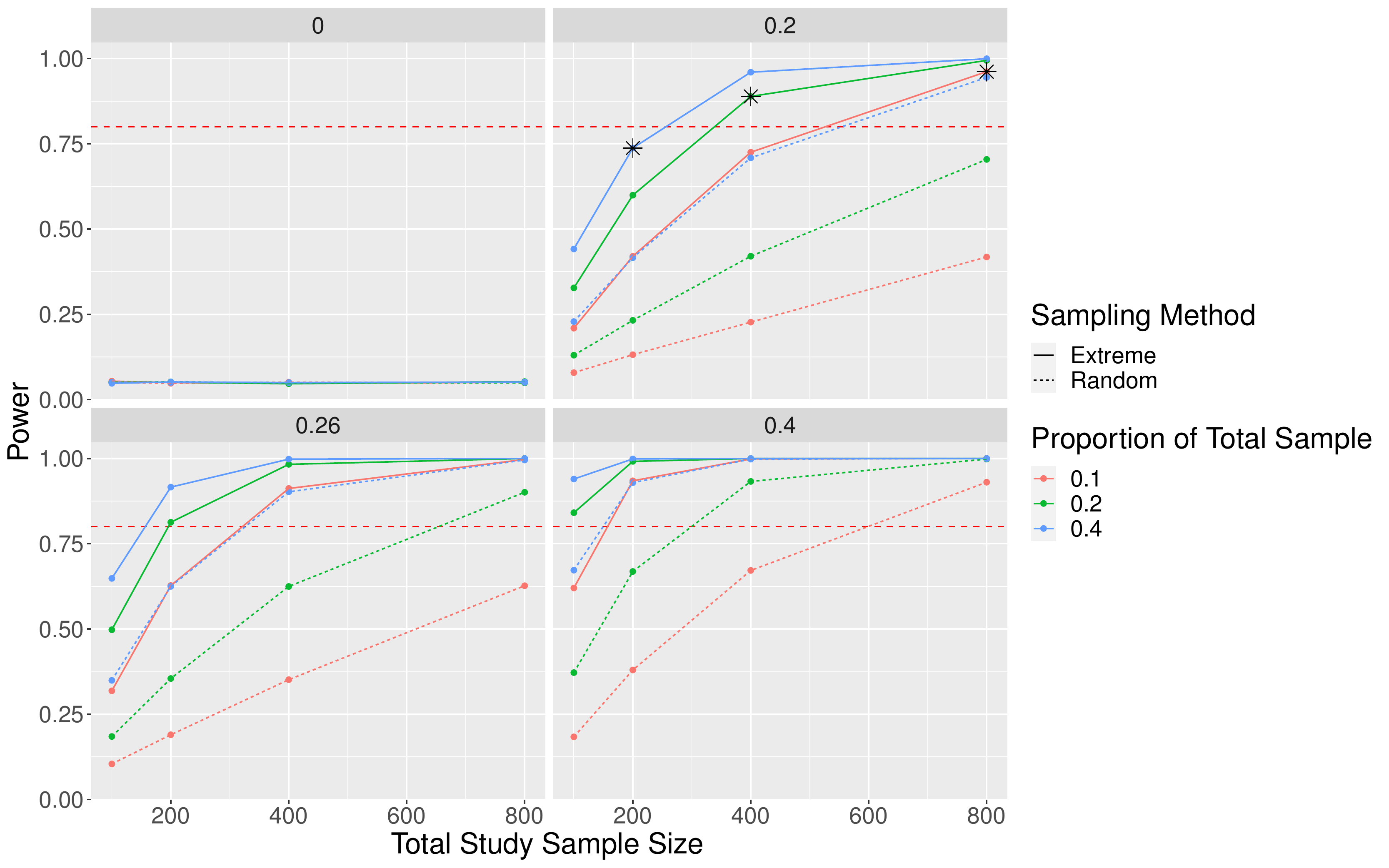}
 \label{fig:Norm_Power}
\end{figure}

In addition to being unbiased and more precise when compared to $\hat{\beta}_{Y,OLS}$, $\hat{\beta}_{Y,ODEB}$ provides a practical increase in power to detect biomarker effect when combined with extreme sampling. First, for ODEB test under all sampling methods, the type I error rate for the test of the null hypothesis $H_0: \beta_Y = 0$ is controlled at the nominal level of 5\% as illustrated in the upper-left panel of Figure \ref{fig:Norm_Power}. When $\beta_Y \ne 0$, random sampling generally has less power to detect a biomarker effect than extreme sampling, with the dashed lines (random sampling) falling below the solid lines (extreme sampling) in each panel. This highlights the utility of focusing on those observations with extreme response values.
Taking $\beta_Y = 0.2$ as an illustrative example of this notion, the stars placed on the upper-right panel emphasize the increase in power afforded by successively more extreme samples.
For the three starred cases, each sample contains 80 biomarker-tested observations.
We can see that a 10\% extreme sample from 800 observations provides 96.2\% power to detect the effect, while a 20\% extreme sample from 400 observations provides 88.9\% power and a 40\% extreme sample from 200 observations provides only 73.8\% power.
A more extreme sample from a larger population has the advantage in terms of power. Intuitively, the more extreme sample contains more information about the relationship, and hence provides a formidable gain in power.

%%%%%%%%%%%%%%%%%%%% ODEB VS. OLS POWER %%%%%%%%%%%%%%%%%%%%%%%%

\begin{figure}
 \centering
  \caption{95\% $\beta_Y$ confidence interval coverage under 20\% of the total study sampled for various $\beta_Y$'s.}
	\includegraphics[width=\textwidth]{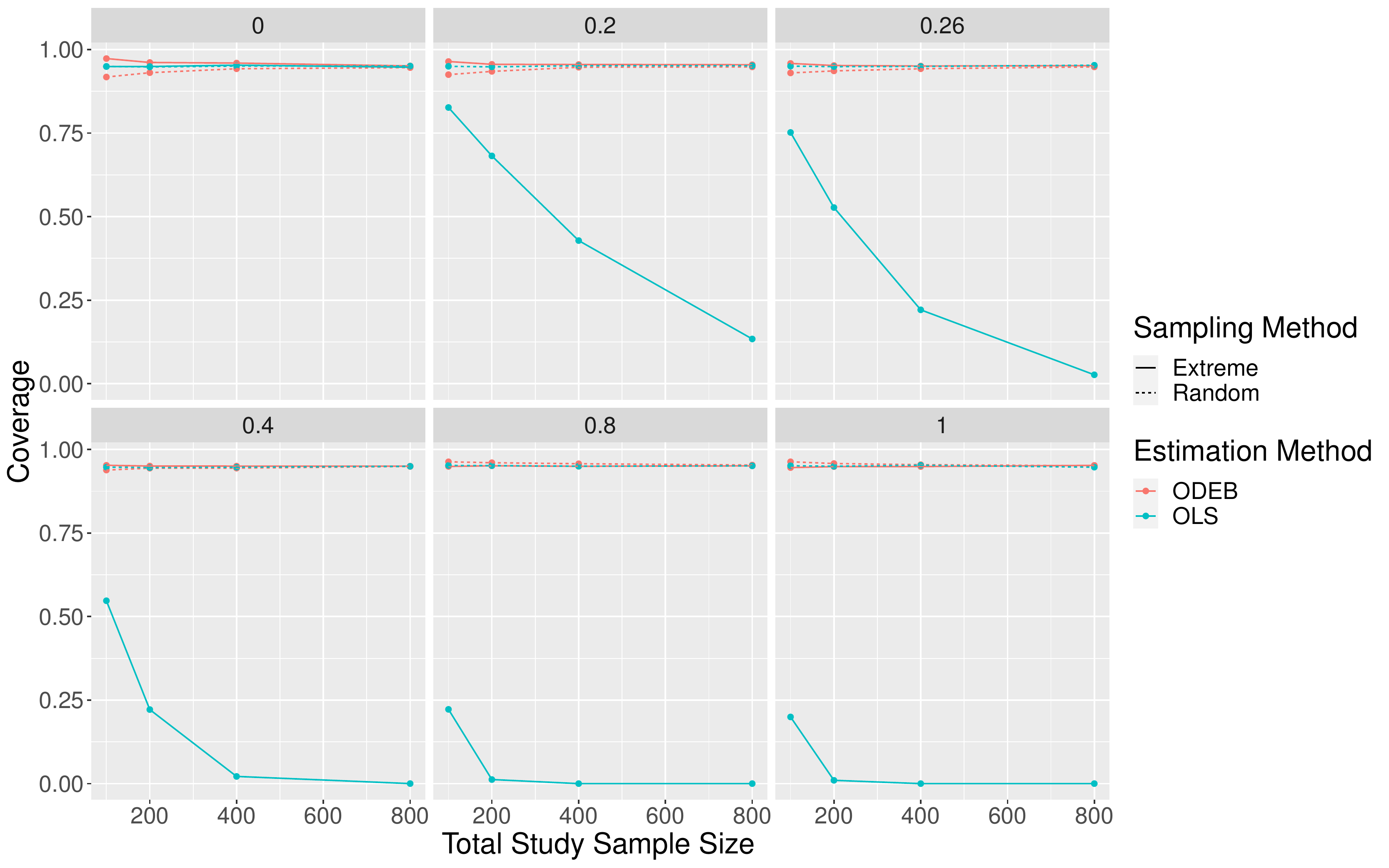}
 \label{fig:Norm_ciCov_20per}
\end{figure}

\begin{figure}
 \centering
  \caption{95\% $\beta_Y$ confidence interval length under 20\% of the total study sampled for various $\beta_Y$'s.}
	\includegraphics[width=\textwidth]{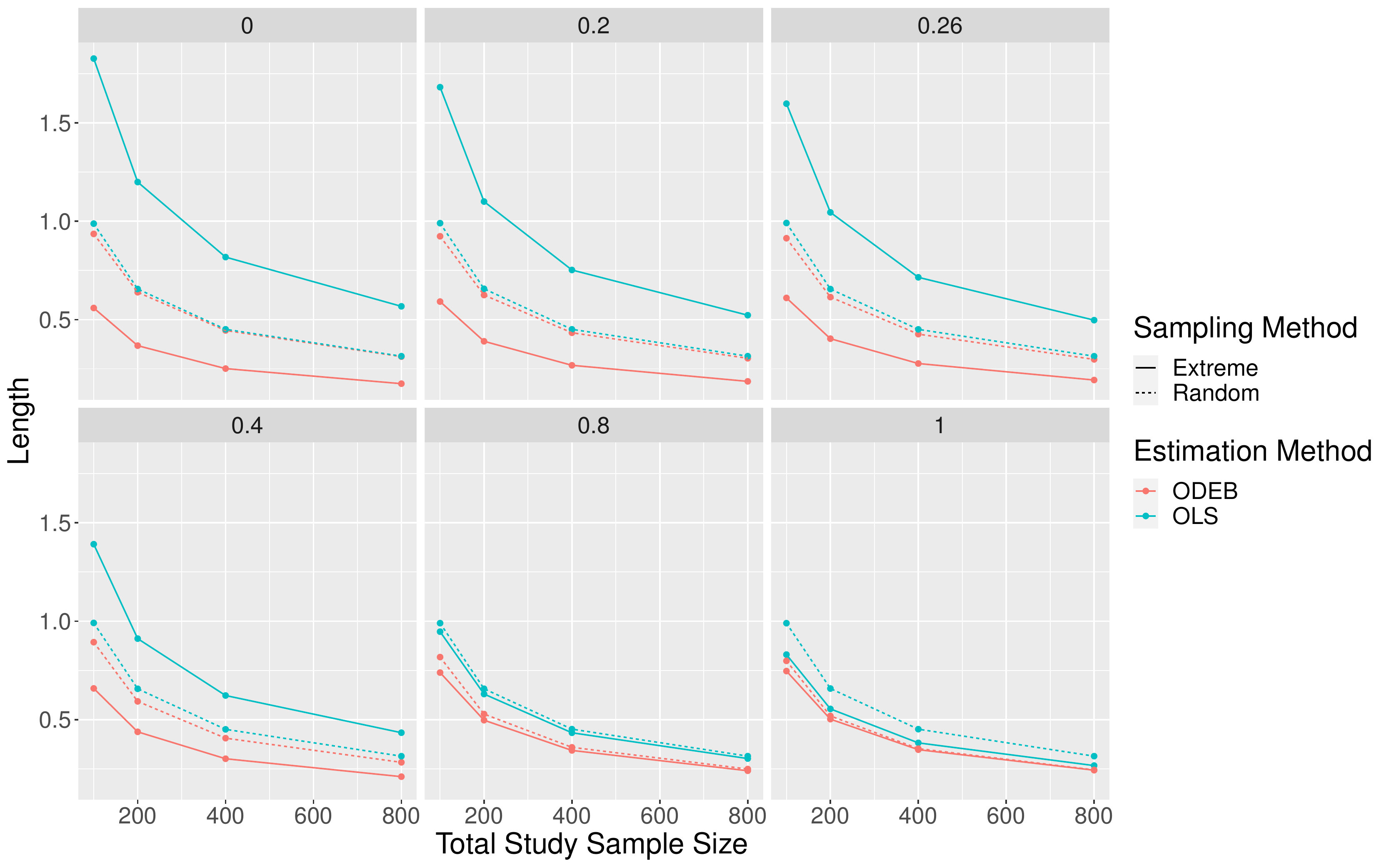}
 \label{fig:Norm_ciLength_20per}
\end{figure}

Figure \ref{fig:Norm_ciCov_20per} studies the performance of the 95\% confidence interval for $\beta_Y$ for OLS and ODEB. The confidence level for the intervals given by ODEB under extreme and random sampling and OLS under random sampling is largely maintained at the nominal 95\%. Unsurprisingly, the OLS-based confidence interval under extreme sampling does not achieve correct nominal 95\% coverage. The poor coverage displayed is worsened both with total study sample size (simply reflecting a larger number of extreme observations contributing to estimation) and the size of $\beta_Y$. This is particularly severe when $\beta_Y=1$, with coverage probabilities of effectively 0\% for $n=200,400, \text{ and } 800$. The OLS-based confidence interval under extreme sampling also displays the longest confidence interval length, though this decreases with $\beta_Y$ to approach the average lengths of the other intervals (Figure \ref{fig:Norm_ciLength_20per}). ODEB estimation affords the most precise confidence interval compared to OLS, particularly when applied to an extreme sample.

%%%%%%%%%%%%%%%%%%% ODEB VS. OLS CI COV %%%%%%%%%%%%%%%%%%%%%%%%

%%%%%%%%%%%%%%%%%%%% ODEB VS. OLS CI LEN %%%%%%%%%%%%%%%%%%%%%%%

\begin{figure}
 \centering
  \caption{Operating characteristics of ODEB and OLS estimation for $\beta_Y=0.2$ under extreme sampling.}
	\includegraphics[width=\textwidth]{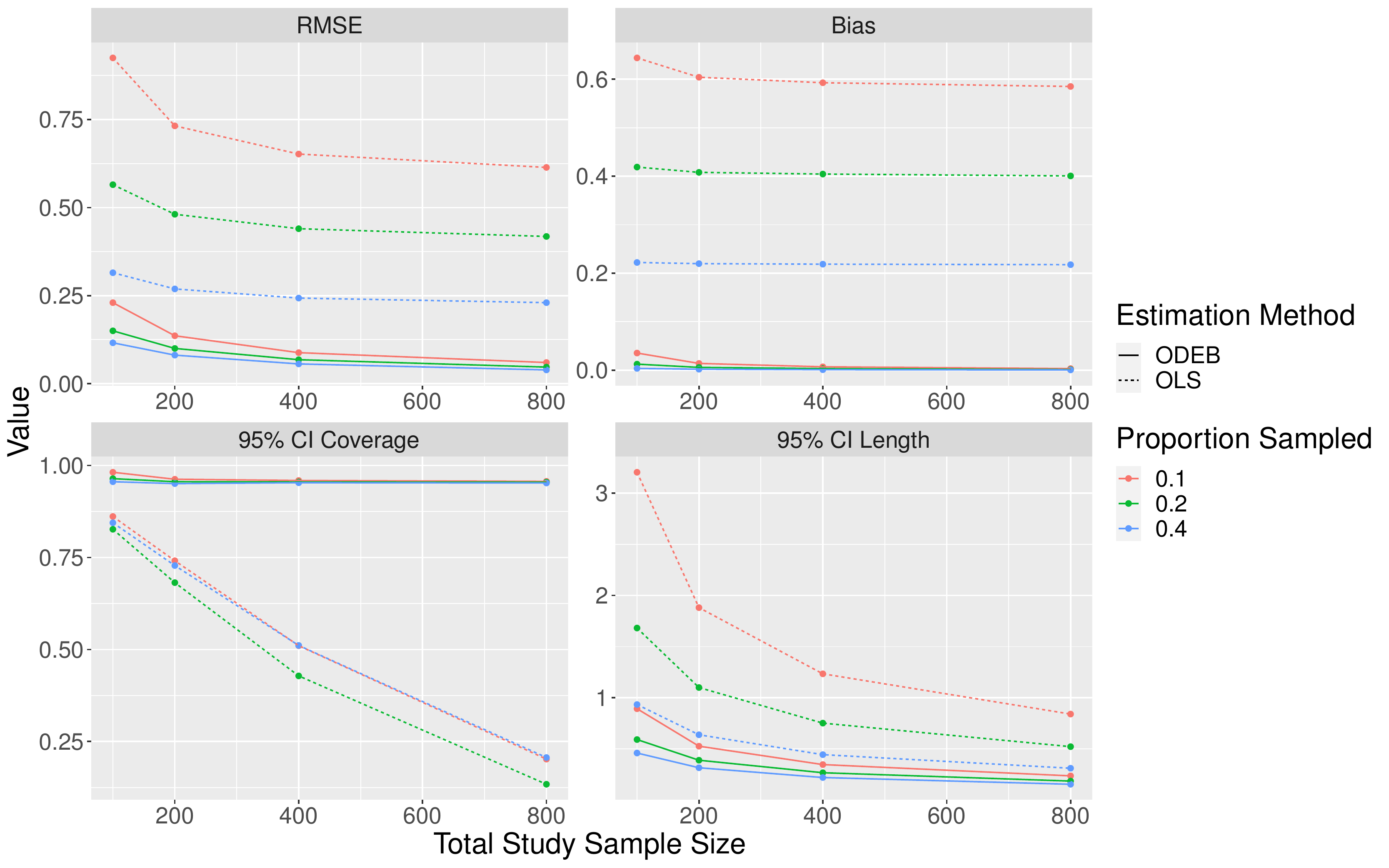}
 \label{fig:singleBeta_ODEB}
\end{figure}

The conclusions of comparison between OLS and our ODEB estimation under example sampling are also summarized in Figure~\ref{fig:singleBeta_ODEB} by fixing $\beta_Y$ and examining the resulting RMSE, bias, average confidence interval coverage, and average confidence interval length for various sampling proportions. RMSE, bias, and confidence interval length are globally higher for OLS estimation compared to ODEB estimation due to the violation of the conditional distribution illustrated in Figure \ref{fig:extremeSampVisual}. Accounting for the sampling scheme, ODEB estimation yields low RMSE and bias. ODEB-based confidence intervals not only maintain nominal coverage, but are shorter in length than those based on $\hat{\beta}_{Y,OLS}$. %In addition, top panels of Figure \ref{fig:extremeSampVisual} shows that, given the total size of the full data, increasing sampling proportion leads to slightly lower RMSE and bias. {I deleted this sentence. It is a trivial conclusion.}

% {\bf More extreme sampling proportions yield slightly higher RMSE and bias (seen in top panels of Figure \ref{fig:singleBeta_ODEB}) at the benefit of increased power.} {\it what does this sentence mean? and how is it shown in the Figure? If you mean the 3 stared cases, we do not show RMSE and bias are higher or not. and it should be "Given same total size of selected sample, sampling a smaller extreme proportion from a bigger full data set yields slightly higher RMSE and bias at the benefit of increased power." Is this an important point or can we just omit it here? } \textcolor{red}{I was referring to the two upper panels of figure 7, where the red line showing 10\% is the highest, followed by 20, then 40. I can definitely clarify this, or if you think it's confusing, we can remove entirely. My figure reference was incorrect, just fixed it.}

%%%%%%%%%%%%%%%%%%% ODEB VS. OLS Fixed B %%%%%%%%%%%%%%%%%%%%%%%

\subsubsection{Simulations with Non-normal Residuals}
To explore the degree to which the ODEB estimation method is affected by violations to the assumption of joint normality, the simulation above was repeated for $\varepsilon_{Y,i}$ following a scaled t distribution (scale $\sqrt{5}$) with $DF \in \{10, 20\}$ and a shifted log-Normal distribution with a mode of zero and variance of 5.
%These error distributions are shown against the reference normal distribution in Figure \ref{fig:regErrors}.

% \begin{figure}
%  \centering
%   \caption{Error distributions used in the Monte Carlo simulations.}
% 	\includegraphics[width=\textwidth]{Final Figures/Error  Visualizations/regErrors_HeavyTail.pdf}
% \label{fig:regErrors}
%\end{figure}

%%%%%%%%%%%%%%%%%%% ODEB Heavy Power %%%%%%%%%%%%%%%%%%%%%%%
\begin{figure}
 \centering
  \caption{Power of $\hat{\beta}_{Y,ODEB}$ under skewed (log-normal) or heavy-tailed (t) residual distributions and an extreme sample of 20\% (Normal residual distribution also plotted for reference).}
	\includegraphics[width=\textwidth]{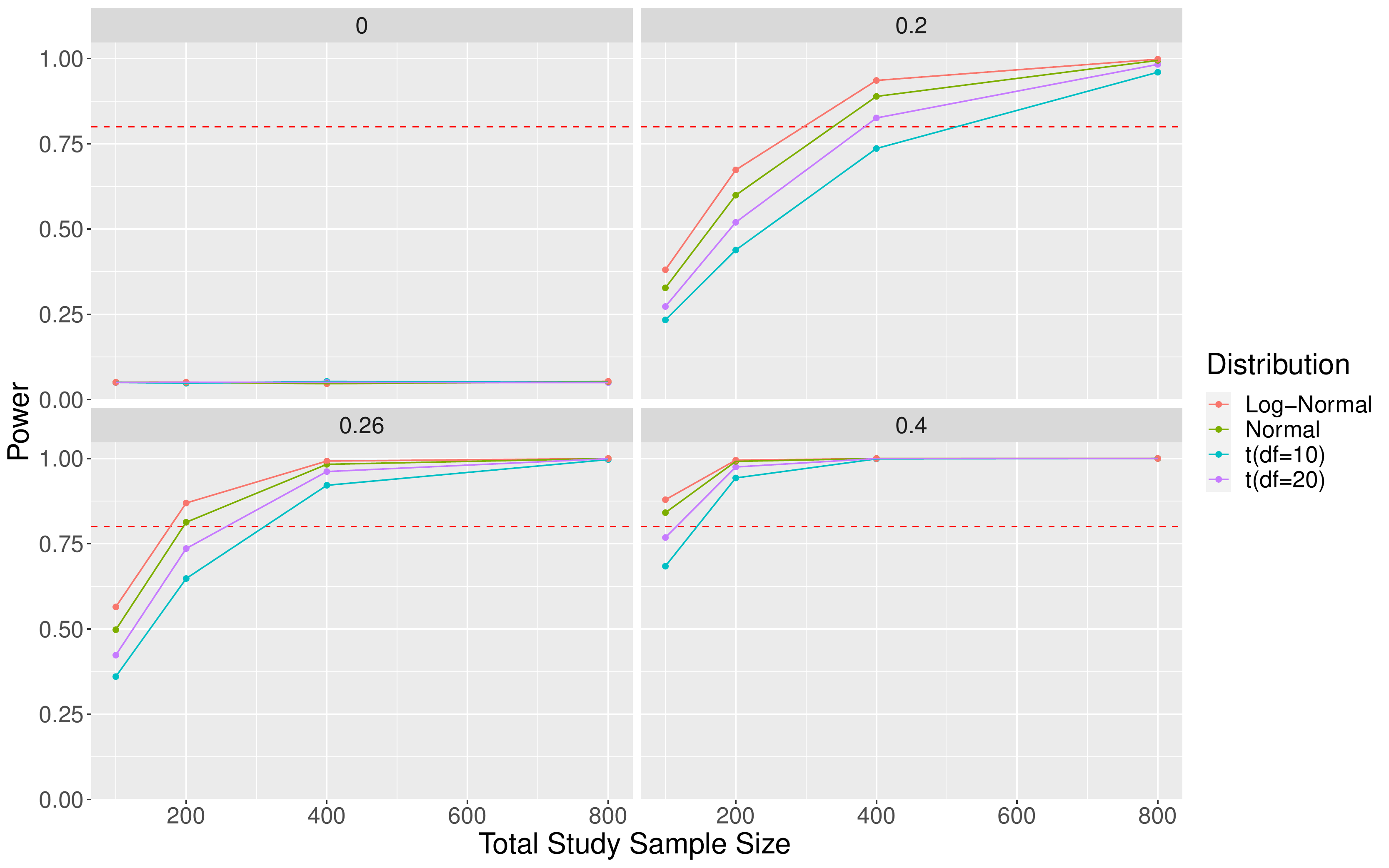}
 \label{fig:heavyPower}
\end{figure}

From the upper-left panel (the case of no biomarker effect with $\beta_Y=0$) of Figure~\ref{fig:heavyPower}, the hypothesis test to detect the effect always has the correct level of $\alpha=0.05$ regardless the normality assumption holds or not. This is expected because, whatever the residual distribution is, there is no effect in the reverse-regression also when the regression effect $\beta_Y=0$. The power of the test improves when the residuals are skewed (shifted log-normal) and deteriorates when the residual distribution becomes more heavily tailed (smaller degrees of freedom for t-distribution). In all simulated settings, the power of detection exceeds $80\%$ (shown as the horizontal dotted red line) when the full sample size is $n_F=800$ (with $n_S=160$ selected for biomarker-testing).

%%%%%%%%%%%%%%%%%%% ODEB Heavy Bias %%%%%%%%%%%%%%%%%%%%%%%
\begin{figure}
 \centering
  \caption{Bias of $\hat{\beta}_{Y,ODEB}$ under skewed (log-normal) or heavy-tailed (t) residual distributions and an extreme sample of 20\% (Normal residual distribution also plotted for reference).}
	\includegraphics[width=\textwidth]{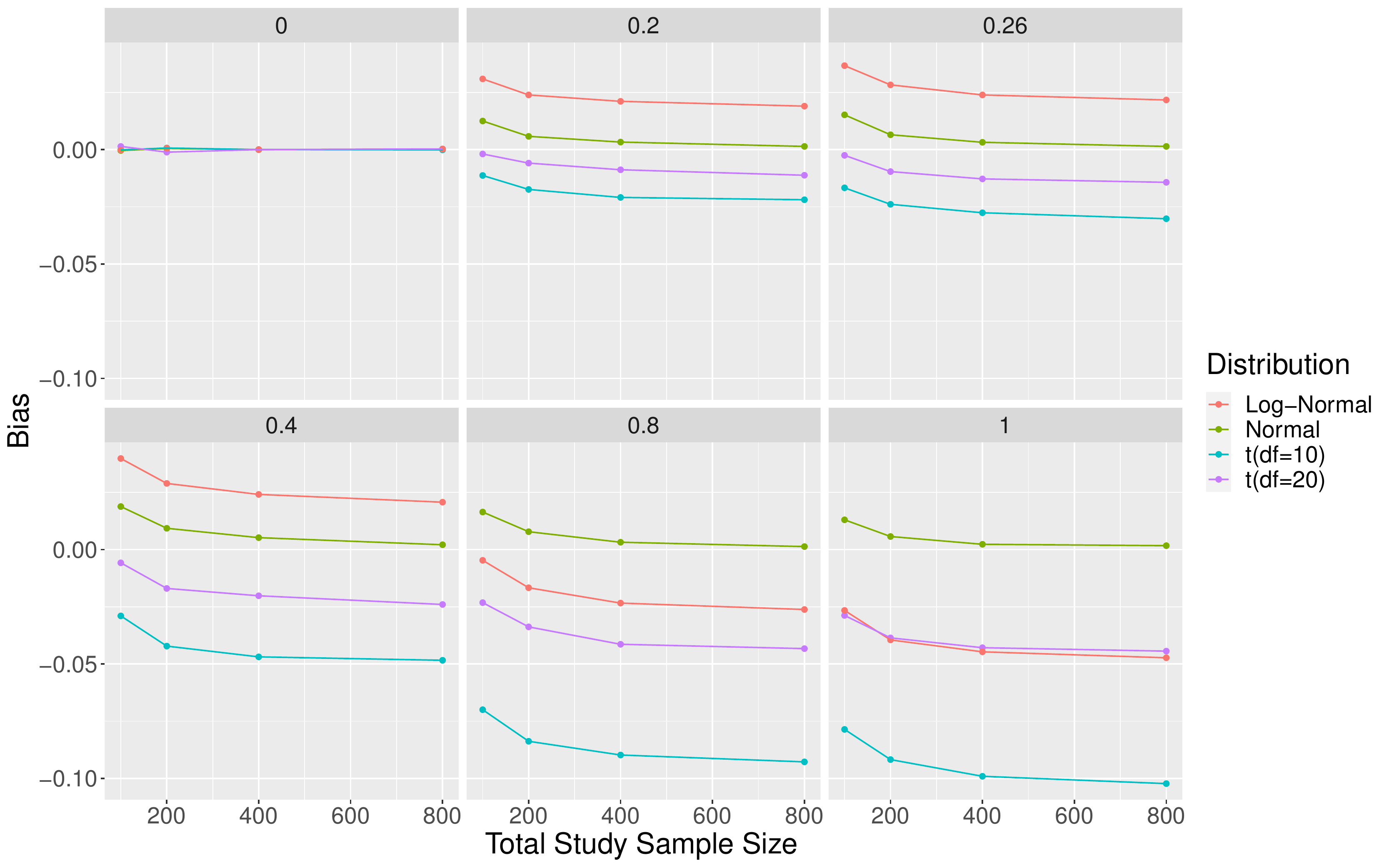}
 \label{fig:heavyBias}
\end{figure}

%%%%%%%%%%%%%%%%%%% ODEB Heavy CI Cov %%%%%%%%%%%%%%%%%%%%%%%
\begin{figure}
 \centering
  \caption{coverage of 95\% confidence interval for $\beta_Y$  under under skewed (log-normal) or heavy-tailed (t) residual distributions and an extreme sample of 20\% (Normal residual distribution also plotted for reference).}
	\includegraphics[width=\textwidth]{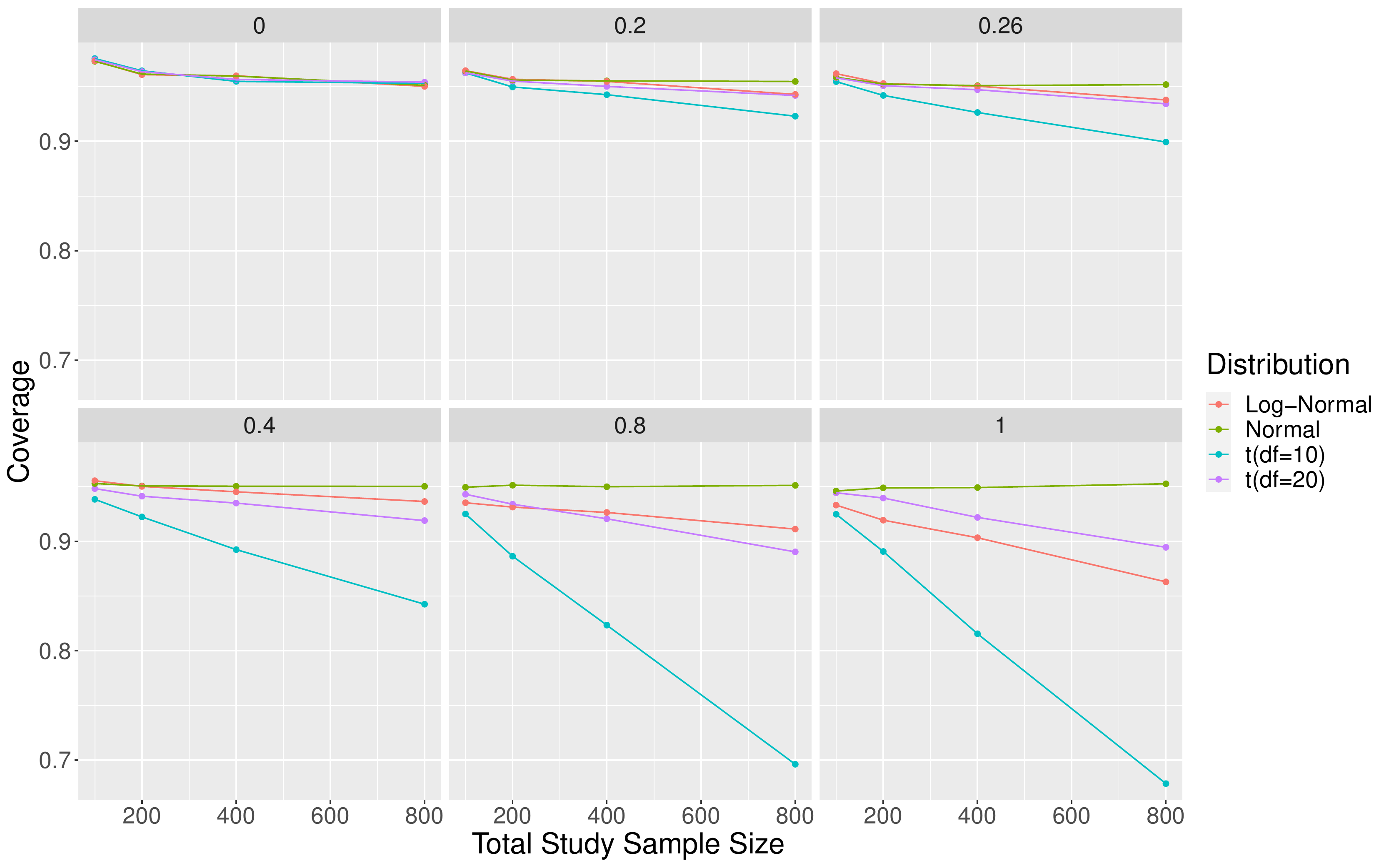}
 \label{fig:heavyCICov}
\end{figure}

The bias of $\hat{\beta}_{Y,ODEB}$ is affected by the residual distributions as seen in Figure~\ref{fig:heavyBias}. Correspondingly, the coverage probability of the $95\%$ confidence interval is no longer valid for non-normal residuals in Figure~\ref{fig:heavyCICov}. The coverage deteriorates when the effect sizes increases, when the sample size increases, and when the tails of residual distribution become heavier.

Overall, when the normality assumption is violated, the effect detection test is still valid and powerful but the effect estimation is no longer reliable.

\subsubsection{Comparison with MSELE Estimator}
Additionally, we compared the ODEB estimator to the MSELE estimator \cite{Zhou2002AOutcome}. Notice that the MSELE estimator cannot be applied directly for data from extreme outcome-dependent sampling. It requires an additional simple random sample (SRS) so that some of the individuals with medial $Y$ values are biomarker-tested also. Thus we conduct the comparison under the ODS experimental design outlined by Zhou et al \cite{Zhou2002AOutcome}.

Again, for each of $B=20,000$ iterations, individual data sets consisting of $(X_1,Y_1)$, ..., $(X_{n_F}, Y_{n_F})$ were randomly generated. This data was eligible for extreme outcome-dependent sampling. An additional SRS of size $n_{SRS}=80$ was generated. For each combination of parameter values described above, an extreme sample of size $\gamma n_F$ was selected. Both $\hat{\beta}_{Y,ODEB}$ and $\hat{\beta}_{Y,MSELE}$ were calculated on the merged sample consisting of the extreme sample and the additional SRS for a total sample size of $n = \gamma n_F + n_{SRS}$. The \textit{ODS} package \cite{odsRPackage} in R was used to calculate $\hat{\beta}_{Y,MSELE}$ and its corresponding standard error.

%%%%%%%%%%%%%%%%%%% MSELE MSE %%%%%%%%%%%%%%%%%%%%%%%
\begin{figure}
 \centering
  \caption{Root mean square error (RMSE) of $\hat{\beta}_{Y,MSELE}$ compared to $\hat{\beta}_{Y,ODEB}$ under 40\% of the total study sampled for various $\beta_Y$'s.}
	\includegraphics[width=\textwidth]{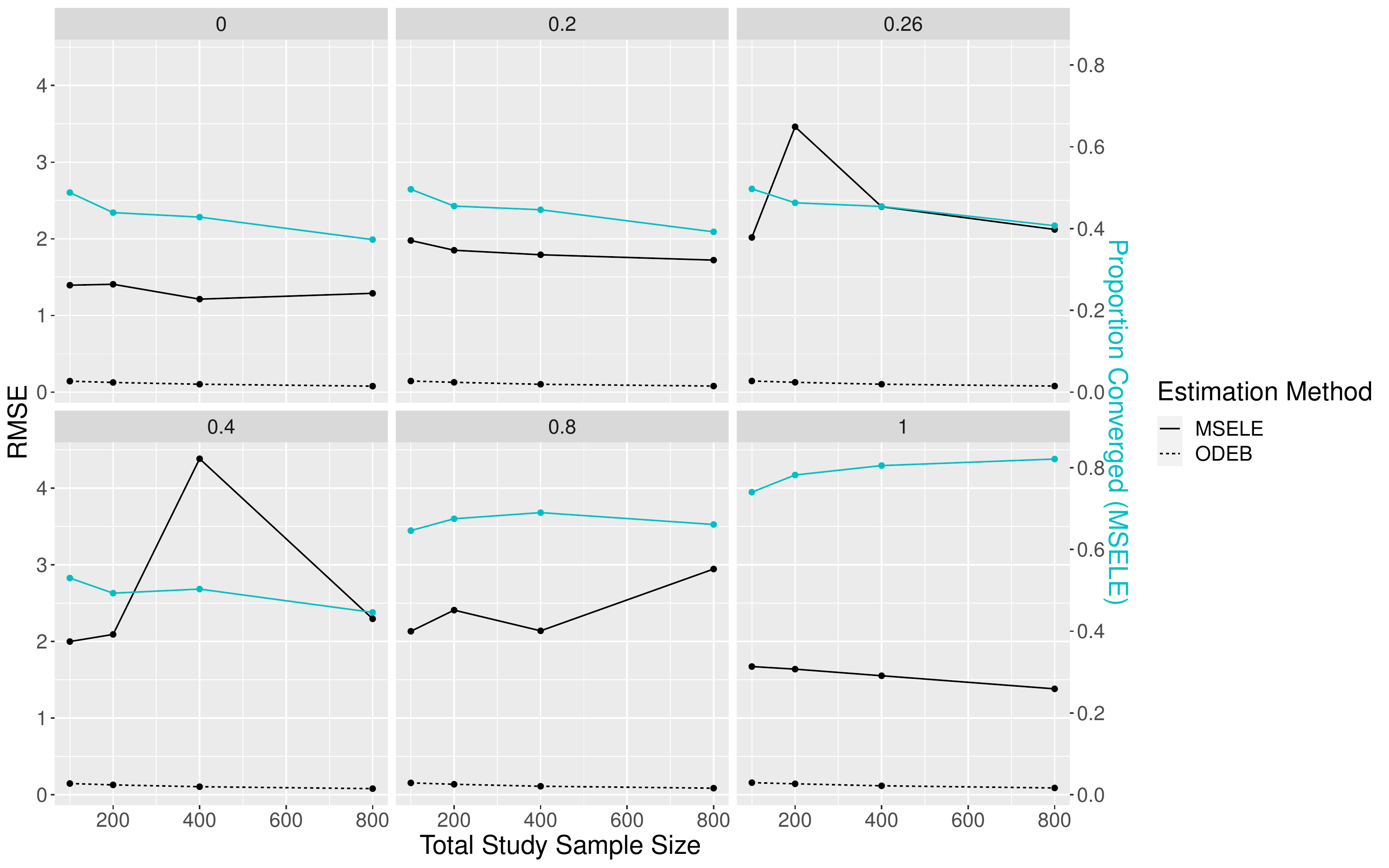}
 \label{fig:mseleRMSE}
\end{figure}

Figure \ref{fig:mseleRMSE} compares the RMSE of $\hat{\beta}_{Y,MSELE}$ and $\hat{\beta}_{Y,ODEB}$.
%$\hat{\beta}_{Y,MSELE}$ is obtained iteratively via Newton-Raphson.
The $\hat{\beta}_{Y,MSELE}$ is obtained from an iterative numerical optimization algorithm, and it does depend on the starting values which poses a challenge in convergence. We also plotted proportion of convergence of $\hat{\beta}_{Y,MSELE}$ out of $B=20,000$ simulation runs on the figure. The convergence rate of $\hat{\beta}_{Y,MSELE}$ is around $50\%$ or lower when the effect size is small. As the effect size increases, the divergence issue alleviates, but there is still a significant proportion of divergence even when $\beta_Y=1$. The RMSE of $\hat{\beta}_{Y,MSELE}$ from the convergent runs are plotted, and are much higher than RMSE of $\hat{\beta}_{Y,ODEB}$ in corresponding settings.

%As such, algorithmic convergence can be a challenge and is dependent on starting values for $\beta$, $\sigma^{2}_{\epsilon_Y}$, and $\pi_i\:(i=1,2,3)$, where $\pi_i$ represents the probability that $Y$ belongs to the upper tail, middle, or lower tail of the domain of $Y$ respectively. Consequently, the proportion converged out of $B=20,000$ iterations is plotted alongside RMSE and is marked on the right side of the figure. Generally, we can see that the RMSE for $\hat{\beta}_{Y,ODEB}$ tends to be lower than that of $\hat{\beta}_{Y,MSELE}$. This was anticipated because $\hat{\beta}_{Y,ODEB}$ uses the information from incomplete observations when estimating the variance of $Y$, while $\hat{\beta}_{Y,MSELE}$ uses only information from those observations with both $Y$ and $X$.

%%%%%%%%%%%%%%%%%%% MSELE MAE %%%%%%%%%%%%%%%%%%%%%%%
\begin{figure}
 \centering
  \caption{Median absolute error (MAE) of $\hat{\beta}_{Y,MSELE}$ compared to $\hat{\beta}_{Y,ODEB}$ under 40\% of the total study sampled for various $\beta_Y$'s.}
	\includegraphics[width=\textwidth]{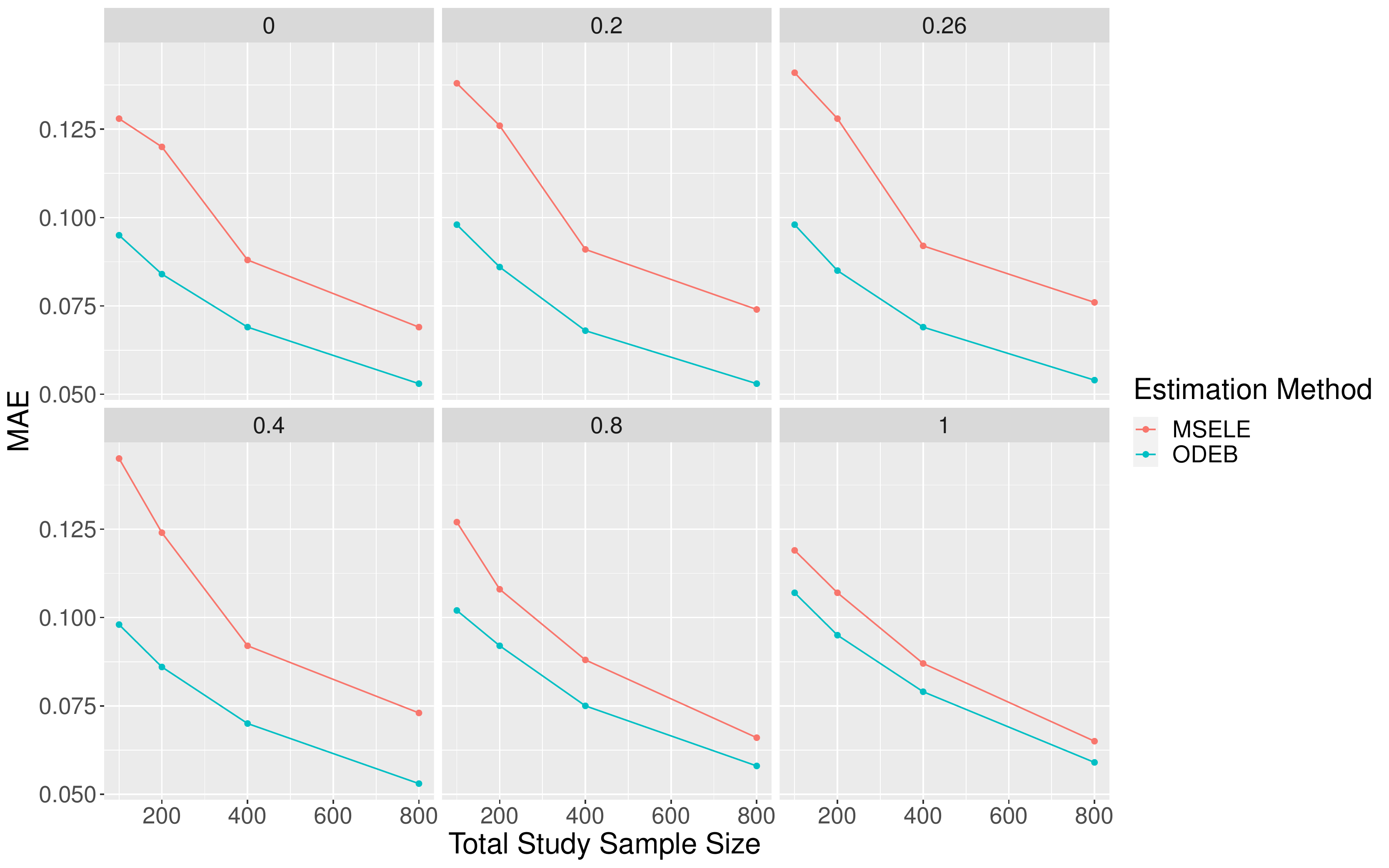}
 \label{fig:mseleMAE}
\end{figure}

The RMSE is heavily affected by outliers. Sometimes, $\hat{\beta}_{Y,MSELE}$ converges to a value that is far from the true value, indicating failure of the iterative algorithm to find the true root. Such wrong convergent cases inflate the RMSE of $\hat{\beta}_{Y,MSELE}$ a lot. To be fairer for $\hat{\beta}_{Y,MSELE}$, Figure \ref{fig:mseleMAE} presents the median absolute error (MAE) comparison instead. The MAE of  $\hat{\beta}_{Y,MSELE}$ is now closer to, but still clearly exceeds, the MAE of $\hat{\beta}_{Y,ODEB}$. This accuracy improvement by $\hat{\beta}_{Y,ODEB}$ is expected here as it utilizes the correct parametric assumption which $\hat{\beta}_{Y,MSELE}$ does not make. As $\beta$ grows larger, this gap in performance between $\hat{\beta}_{Y,MSELE}$ and $\hat{\beta}_{Y,ODEB}$ shrinks.
%Occasionally, the iterative procedure met the computational convergence criteria but provided an estimate of $\beta$ which was a very large outlier, indicating failure of the algorithm to find the true root. This resulted in the unusual trends in RMSE within the MSELE estimator in Figure \ref{fig:mseleRMSE}. For a more fair visualization, Figure \ref{fig:mseleMAE} shows instead the median absolute error (MAE). Intuitively, very similar performance of the two estimators is seen across true $\beta$'s and study sample sizes, with ODEB again showing the benefit of using the information from incomplete observations. As $\beta$ grows larger, this gap in performance between $\hat{\beta}_{Y,MSELE}$ and $\hat{\beta}_{Y,ODEB}$ shrinks.

\subsection{Data Analysis/Application to BISP2 Trial}
\label{s:analysis}

The dependent variable of interest in the BISP2 Trial, pain intensity, is measured via the BPI to represent an underlying pain continuum for all participants. A pilot sample of 31 participants in the high and low tails of worst pain experienced at 48-hours post shoulder pain induction was selected for initial biomarker-testing. As the BPI is a discrete index represented on a continuous scale, there were ties for the highest and lowest pain scores. To determine which of the equal scores were included in the pilot sample, those who responded vs. did not respond to the treatment were selected in a balanced manner. Based on promising results from the initial pilot sample, a follow-up sample of 57 participants was selected in the same manner. All biomarkers were logarithm (base 10, for interpretability) transformed based on prior knowledge of the typical biomarker distribution. We model the relationship between pain and log-transformed biomarker linearly as:

$$y_{Pain} = \alpha + \beta log(x_{Biomarker}) + \epsilon$$

\begin{figure}
 \centering
  \caption{Worst pain experienced at 48-hours shoulder pain induction for each sampling stage, combined stages, and the full BISP2 study population.}
	\includegraphics[width=\textwidth]{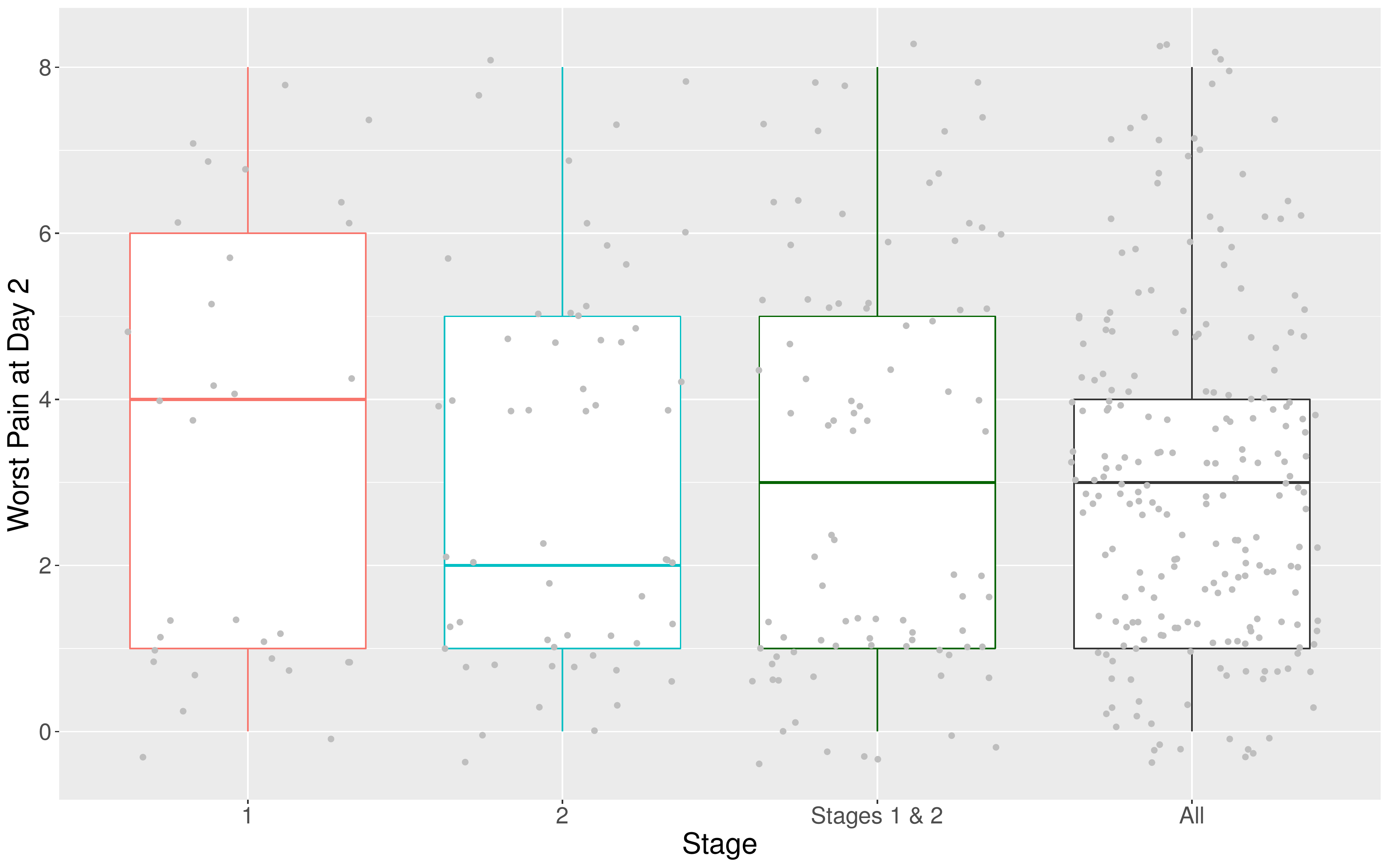}
 \label{fig:wpDistBISP2}
\end{figure}

Two pain-related outcomes of interest were explored using the above methodology: worst pain experienced at 48-hours post shoulder pain induction and change in worst pain from 48- to 96-hours post induction. Additionally, a self-reported disability score (Quick-DASH) at 48-hours post induction and change in Quick-DASH score from 48- to 96-hours post induction was investigated as outcomes to assess endpoint specificity. The results of modeling the association between pain-related outcomes and log-transformed biomarkers are discussed below. The distribution of worst pain at 48-hours in each stage of sampling is given in Figure \ref{fig:wpDistBISP2}. For each pain-related outcome, reverse-regression based estimates of the effect of each log-transformed biomarker on pain ($\beta$) are reported with standard errors, 95\% confidence intervals, and p-values. The the reverse-regression model assumption was assessed using diagnostic plots of residuals.

The results of the analysis of the combined stages 1 and 2 data for worst pain and change in pain are shown in Tables \ref{table:wpResBISP2} and \ref{table:cpResBISP2}. Table \ref{table:wpResBISP2} shows evidence of association between CCL2 and worst pain at 48 hours, with an effect estimate of 1.92 (95\% CI: 0.24, 3.61). This indicates that for a 10-fold increase in CCL2, we expect the worst pain at 48 hours to increase by 1.92 points on the BPI scale. For change in pain, there is evidence of association from CRP ($\hat\beta = 0.71$ (95\% CI: 0.05, 1.37)). We also see potential from IL10 and IL6, however the evidence in this sample is insufficient to conclude such.
In additional analyses for self-reported disability outcomes (whose full result tables are omitted for succinctness), CRP shows evidence of association with Quick-DASH score. For a 10-fold increase in CRP, we expect an increase of 3.92 points in the Quick-DASH score. %{\it where in the table are the last two results? do we omitted some details? need to state so.} \textcolor{red}{Sam and I had discussed previously that it was overkill to include two additional tables with the entire analysis for quick DASH, but I still thought it was interesting from an endpoint specificity point of view that CRP was found to have an association with q dash. I clarified above that they are omitted results, let me know if this seems clear enough for the reader.}

%%%%%%%%%%%%%%% Worst Pain Estimation Results%%%%%%%%%%%%%%%%%%%
\begin{table}
\centering
\caption{Summaries from univariable reverse-regression based extreme sampling estimation and inference: Worst pain at 48 hours regressed on log-scale biomarkers.}
\scalebox{0.9}{
\begin{tabular}{lccccc}
\toprule
                      & Estimate & Std. Error & LCL    & UCL   & P-Value  \\
\midrule
% \rowcolor[HTML]{FFFF00}
\rowcolor[HTML]{D0CECE}
log(CCL2)             & 1.923    & 0.847      & 0.238  & 3.607 & 0.026                       \\
log(CRP)              & 0.369    & 0.296      & -0.220 & 0.958 & 0.216                       \\
log(CXCL6)            & -0.775   & 0.973      & -2.710 & 1.161 & 0.428                       \\
log(IL17/IL17A)       & 0.682    & 1.059      & -1.423 & 2.787 & 0.521                       \\
log(IL10)             & 0.742    & 1.290      & -1.823 & 3.306 & 0.567                       \\
log(TNF Alpha)        & 0.690    & 1.403      & -2.099 & 3.479 & 0.624                       \\
log(BDNF)             & -0.139   & 0.313      & -0.760 & 0.483 & 0.658                       \\
log(IL6)              & 0.376    & 1.475      & -2.556 & 3.307 & 0.799                       \\
log(Beta NGF)         & 0.323    & 1.823      & -3.301 & 3.947 & 0.860                       \\
log(TNF R1)           & 0.269    & 1.641      & -2.993 & 3.531 & 0.870                       \\
log(Oncostatin M OSM) & 0.176    & 1.737      & -3.278 & 3.629 & 0.920                       \\
log(Substance P)      & 0.107    & 1.085      & -2.05  & 2.265 & 0.921                       \\
log(Cortisol)         & 0.020    & 0.491      & -0.955 & 0.995 & 0.968                       \\
\bottomrule
\end{tabular}
}
\label{table:wpResBISP2}
\end{table}

%%%%%%%%%%%%%%%%Change in Pain Estimation Results%%%%%%%%%%%%%%%%%%%
\begin{table}[]
\centering
\caption{Summaries from univariable reverse-regression based extreme sampling estimation and inference: Change in worst pain
from 48 hours to 96 hours regressed on log-scale biomarkers.}
\scalebox{0.9}{
\begin{tabular}{lccccc}
\toprule
                      & Estimate & Std. Error & LCL    & UCL   & P-Value \\
\midrule
% \rowcolor[HTML]{FFFF00}
\rowcolor[HTML]{D0CECE}
log(CRP)              & 0.710     & 0.332      & 0.050   & 1.370  & 0.039         \\
\rowcolor[HTML]{D0CECE}
log(IL10)             & 2.752    & 1.452      & -0.135 & 5.639 & 0.065            \\
\rowcolor[HTML]{D0CECE}
log(IL6)              & 3.078    & 1.661      & -0.225 & 6.380  & 0.071           \\
log(TNF Alpha)        & 2.269    & 1.599      & -0.910  & 5.448 & 0.163           \\
log(Oncostatin M OSM) & 2.262    & 1.991      & -1.697 & 6.22  & 0.262            \\
log(TNFR1)            & 2.018    & 1.883      & -1.726 & 5.761 & 0.289            \\
log(Beta NGF)         & 1.456    & 2.104      & -2.727 & 5.639 & 0.492            \\
log(CCL2)             & 0.659    & 0.988      & -1.305 & 2.623 & 0.507            \\
log(Cortisol)         & -0.376   & 0.566      & -1.502 & 0.750  & 0.509           \\
log(Substance P)      & 0.411    & 1.256      & -2.086 & 2.909 & 0.744            \\
log(BDNF)             & -0.083   & 0.362      & -0.803 & 0.637 & 0.819            \\
log(IL17/IL17A)       & 0.174    & 1.228      & -2.267 & 2.615 & 0.888            \\
log(CXCL6)            & -0.062   & 1.130      & -2.308 & 2.183 & 0.956            \\
\bottomrule
\end{tabular}}
\label{table:cpResBISP2}
\end{table}

%%%%%%%%%%%%%%% Model Checking Example %%%%%%%%%%%%%%%%%%%
\begin{figure}
 \centering
  \caption{Model checking: Normal QQ plot example for the change in pain outcome and the reverse regression of $Y_{Pain\:Change}$ on $log(X_{CRP})$.}
	\includegraphics[width=\textwidth]{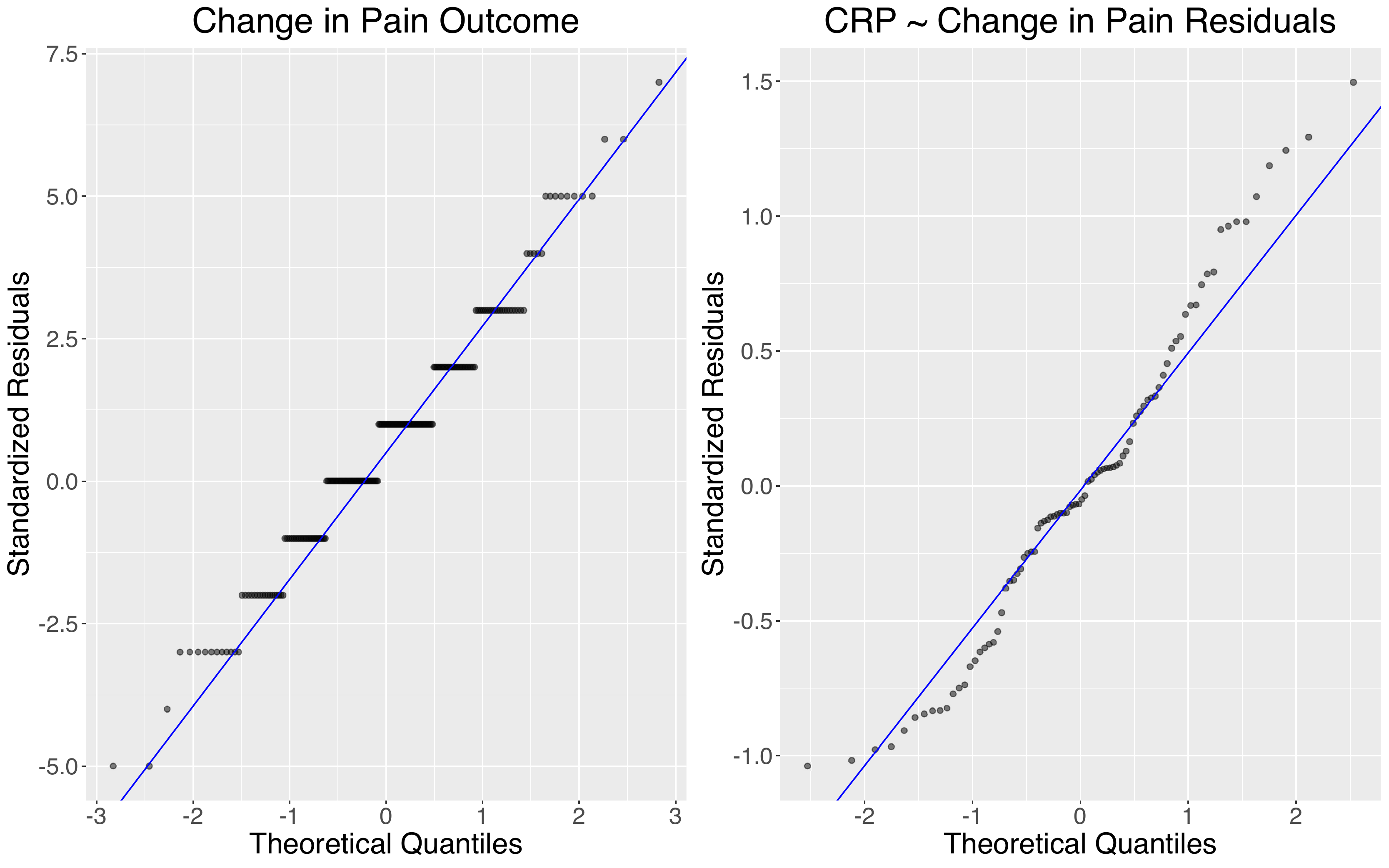}
 \label{fig:qqplots}
\end{figure}

Figure \ref{fig:qqplots} provides the model checking diagnostic plots for the change in pain outcome and the residuals of the reverse regression of log-transformed CRP on change in pain. Our joint normality assumption appears to be reasonable for this example.

\section{Discussion}
\label{s:discuss}

We proposed a new analysis method for the extreme outcome-dependent sampling based on the joint normality assumption. Traditionally, the outcome-dependent sampling data is analyzed through likelihood approach~\cite{Zhou2002AOutcome} with an additional SRS to provide some observations in the middle range of response values. The main advantage of the likelihood approach is that no parametric distributional assumption is needed. In contrast, through the joint normality assumption, our method provides two significant practical advantages. First, our method allows concentration of all sampling to the more informative individuals with the most extreme response values, resulting in great cost reduction for the expansive biomarker-testing. Second, our method is simple and can be conducted with existing standard statistical software. As seen in the simulation, often the likelihood method diverges or it converges to wrong parameter values, thus its data analysis often requires much care from an expert statistician. On the other hand, our method only requires applications of simple formulas on standard least-square estimates from the reverse-regression, and can be handled by any practitioner with a minimum statistical training. Furthermore, our normality assumption can be checked using some standard model checking techniques. When the normality assumption is violated, the confidence interval coverage deteriorate only for very heavy-tailed residuals, but the hypothesis testing for effect detection always remain valid. Thus our new EODS analysis method is particularly suited for exploratory biomarker studies due to its cost savings and ease of usage.

This paper is a first work demonstrating the feasibility of EODS regression analysis assuming bi-variate joint normal distribution of a response variable and a biomarker variable. In the future, we will extend these analysis to more applications. For the multiple regression analysis, in principle we can similarly derive formulas assuming joint normal distribution of a response variable and multiple biomarker variables. To allow analysis of longitudinal data, we need to consider analysis from joint distribution of multiple responses (over various time points) and biomarkers.

%References
\bibliographystyle{vancouver}
\bibliography{references}
%No line space in code before special sections or it will indent the headings
%\vspace{\baselineskip} \\

%End of paper special sections
%Acknowledgements
%\textbf{Acknowledgements}

%\vspace{\baselineskip}\\

%Competing interests
%\textbf{Declaration of conflicting interests}
%\vspace{\baselineskip}\\

%funding
\textbf{Funding}
Dr. Wu was partially supported by the funding from the National Institutes of Health and National Institute of Arthritis and Musculoskeletal and Skin Diseases (AR055899). All authors were
independent from this funding source.

%\vspace{\baselineskip}\\
%supplementary material
%\textbf{Supplementary material}\\
%i.e. code/package/shiny app and where to find

% Appendix Sections
\appendix
\setcounter{secnumdepth}{0}

\section{Appendices}

\subsection{A.1. Mathematical derivation of formulas relating parameters of regression and reverse-regression}

We assume that $(X,Y)$ follows the bivariate normal distribution
$$
N\left( \left( \begin{array}{c} \mu_X \\ \mu_Y \end{array} \right), \left( \begin{array}{cc} \sigma^2_X & \rho \sigma_X \sigma_Y \\ \rho \sigma_X \sigma_Y & \sigma_Y^2 \end{array} \right) \right).
$$
Under this assumption, we consider how to get regression parameters $\alpha_Y$, $\beta_Y$ and $\sigma_{\varepsilon_Y}$ from the reverse-regression parameters $\alpha_X$, $\beta_X$ and $\sigma_{\varepsilon_X}$.

Conditional on $X$, we have the regression equation
\begin{equation}\label{eq:YregX}
    Y = \alpha_Y + \beta_Y X +  \varepsilon_Y,
\end{equation}
where $\varepsilon_Y \sim N(0, \sigma^2_{\varepsilon_Y})$ with $\sigma^2_{\varepsilon_Y}= (1 - \rho^2) \sigma^2_Y$.
Similarly, conditional on $Y$, we have the reverse-regression equation
\begin{equation}\label{eq:XregY}
    X = \alpha_X + \beta_X Y + \varepsilon_X,
\end{equation}
where $\varepsilon_X \sim N(0, \sigma^2_{\varepsilon_X})$ with $\sigma^2_{\varepsilon_X}= (1 - \rho^2) \sigma^2_X$.

%Notice that, for the data selected under extreme sampling based on $Y$ values, the regression model~\eqref{eq:YregX} no longer holds, but the regression model~\eqref{eq:XregY} still holds.

Firstly we derive how the regression parameters of equation~\eqref{eq:XregY} relate to the bivariate normal distribution parameters. To do this, we consider the standardized versions of $X$ and $Y$ as $X^*=\frac{X- \mu_X}{\sigma_X}$ and $Y^*=\frac{Y- \mu_Y}{\sigma_Y}$. Then clearly, $(X^*,Y^*)$ follows the bivariate normal distribution
$$
N\left( \left( \begin{array}{c} 0 \\ 0 \end{array} \right), \left( \begin{array}{cc} 1 & \rho  \\ \rho  & 1 \end{array} \right) \right).
$$
Hence we have
\begin{equation}\label{eq:X*regY*}
    X^* = 0 + \rho Y^* + \sqrt{1-\rho^2} Z,
\end{equation}
where $Z \sim N(0,1)$ is independent of $Y$. Plug $X^*=\frac{X- \mu_X}{\sigma_X}$ and $Y^*=\frac{Y- \mu_Y}{\sigma_Y}$ into equation~\eqref{eq:X*regY*} and compare with equation~\eqref{eq:XregY}, we can express the reverse-regression parameters in terms of the bivariate normal distribution parameters as
\begin{equation}\label{eq:Para}
\begin{array}{cl}
     \alpha_X &  =\mu_X - \beta_X \mu_Y \qquad = \mu_X - \rho \frac{\sigma_X}{\sigma_Y} \mu_Y,\\
     \beta_X & = \rho \frac{\sigma_X}{\sigma_Y}, \\
     \sigma^2_{\varepsilon_X} & = (1 - \rho^2) \sigma^2_X.
\end{array}
\end{equation}

Secondly, by symmetry, similar derivations give the regression parameters' formulas in terms of the bivariate normal distribution parameters as
\begin{equation}\label{eq:ParaY}
\begin{array}{cl}
     \alpha_Y &  =\mu_Y - \beta_Y \mu_X \qquad = \mu_Y - \rho \frac{\sigma_Y}{\sigma_X} \mu_X,\\
     \beta_Y & = \rho \frac{\sigma_Y}{\sigma_X}, \\
     \sigma^2_{\varepsilon_Y} & = (1 - \rho^2) \sigma^2_Y.
\end{array}
\end{equation}

We now find formulas to express regression parameters $\alpha_Y$, $\beta_Y$ and $\sigma_{\varepsilon_Y}$ in terms of $\alpha_X$, $\beta_X$, $\sigma_{\varepsilon_X}$, $\mu_Y$ and $\sigma_Y$.

We start from the parameter of most interest $\beta_Y$. Combine the middle equations in \eqref{eq:Para} and \eqref{eq:ParaY}, we get
\begin{equation}\label{eq:A2}
\beta_Y  = \rho \frac{\sigma_Y}{\sigma_X} = (\rho \frac{\sigma_X}{\sigma_Y}) \frac{\sigma^2_Y}{\sigma^2_X}= \beta_X \frac{\sigma^2_Y}{\sigma^2_X}.
\end{equation}
Now, to get the desired expression of $\beta_Y$, we only need to express $\sigma^2_X$ in terms of $\alpha_X$, $\beta_X$, $\sigma_{\varepsilon_X}$, $\mu_Y$ and $\sigma_Y$.

From the middle equation in~\eqref{eq:Para},
\begin{equation}\label{eq:A1}
\beta_X = \rho \frac{\sigma_X}{\sigma_Y} \qquad  \Rightarrow \qquad \beta_X {\sigma_Y} = \rho {\sigma_X}.
\end{equation}
Square both sides of equation~\eqref{eq:A1} and divides the last equation in \eqref{eq:Para}, we get
$$
\frac{\beta^2_X {\sigma^2_Y}}{\sigma^2_{\varepsilon_X}} = \frac{\rho^2}{1-\rho^2}.
$$
Therefore, we solve $\rho^2$ in this expression to get
\begin{equation}\label{eq:rho2}
\rho^2 = \frac{1}{1+\frac{\sigma^2_{\varepsilon_X}}{\beta^2_X {\sigma^2_Y}}}, \qquad \mbox{and} \qquad 1-\rho^2 = 1- \frac{1}{1+\frac{\sigma^2_{\varepsilon_X}}{\beta^2_X {\sigma^2_Y}}}= \frac{\frac{\sigma^2_{\varepsilon_X}}{\beta^2_X {\sigma^2_Y}}}{1+\frac{\sigma^2_{\varepsilon_X}}{\beta^2_X {\sigma^2_Y}}} = \frac{1}{1+\frac{\beta^2_X {\sigma^2_Y}}{\sigma^2_{\varepsilon_X}}}.
\end{equation}
Plug this back into the last equation in \eqref{eq:Para} to solve for $\sigma^2_X$, we get
$$
\sigma^2_X \qquad = \frac{\sigma^2_{\varepsilon_X}}{1-\rho^2} \qquad = \sigma^2_{\varepsilon_X}[1+\frac{\beta^2_X {\sigma^2_Y}}{\sigma^2_{\varepsilon_X}}] \qquad = \sigma^2_{\varepsilon_X} + \beta^2_X {\sigma^2_Y}.
$$

Put this into equation~\eqref{eq:A2}, we have the desired expression for $\beta_Y$ as
\begin{equation}\label{eq:BetaY_BetaX}
\beta_Y  \qquad = \beta_X \frac{\sigma^2_Y}{\sigma^2_{\varepsilon_X} + \beta^2_X {\sigma^2_Y}} \qquad = \frac{1}{\frac{\sigma^2_{\varepsilon_X}}{\sigma^2_Y} + \beta^2_X } \beta_X \qquad = \frac{\beta_X{\sigma^2_Y}}{{\sigma^2_{\varepsilon_X}} + \beta^2_X {\sigma^2_Y} }.
\end{equation}

Next, we express $\alpha_Y$ in terms of $\alpha_X$, $\beta_X$, $\sigma_{\varepsilon_X}$, $\mu_Y$ and $\sigma_Y$. From the first equation in~\eqref{eq:Para}, $\mu_X = \alpha_X + \beta_X \mu_Y$. Plug this and \eqref{eq:BetaY_BetaX} both into the first equation of~\eqref{eq:ParaY}, we get
\begin{equation}\label{eq:alphaY_BetaX}
 \alpha_Y   \qquad =\mu_Y - \beta_Y \mu_X \qquad  =\mu_Y - (\frac{\beta_X{\sigma^2_Y}}{{\sigma^2_{\varepsilon_X}} + \beta^2_X {\sigma^2_Y} })  (\alpha_X + \beta_X \mu_Y) \qquad = \frac{\sigma^2_{\varepsilon_X}\mu_Y - \alpha_X  \beta_X{\sigma^2_Y}}{{\sigma^2_{\varepsilon_X}} + \beta^2_X {\sigma^2_Y} }.
\end{equation}

Finally, for the expression of $\sigma^2_{\varepsilon_Y}$, we plug \eqref{eq:rho2} into the last equation of~\eqref{eq:ParaY} to get
\begin{equation}\label{eq:sigmaEY_BetaX}
 \sigma^2_{\varepsilon_Y}   \qquad =\frac{1}{1+\frac{\beta^2_X {\sigma^2_Y}}{\sigma^2_{\varepsilon_X}}} \sigma^2_Y \qquad = \frac{\sigma^2_Y \sigma^2_{\varepsilon_X}}{\sigma^2_{\varepsilon_X}+\beta^2_X {\sigma^2_Y} }.
\end{equation}

Furthermore, we also derive the variance formula for the point estimator of $\beta_Y$. From the equation~\eqref{eq:BetaY_BetaX}, we find the following partial derivatives.
\begin{equation}\label{eq:BetaY_PartDeri}
\begin{array}{cll}
\frac{\partial}{\partial \beta_X}\beta_Y  & = \frac{1}{\frac{\sigma^2_{\varepsilon_X}}{\sigma^2_Y} + \beta^2_X } + \beta_X \frac{-1}{(\frac{\sigma^2_{\varepsilon_X}}{\sigma^2_Y} + \beta^2_X )^2} (2 \beta_X)  & = \frac{\frac{\sigma^2_{\varepsilon_X}}{\sigma^2_Y}-\beta^2_X}{(\frac{\sigma^2_{\varepsilon_X}}{\sigma^2_Y} + \beta^2_X )^2},  \\
\frac{\partial}{\partial (\sigma^2_{\varepsilon_X})}\beta_Y  & = \frac{-\beta_X}{(\frac{\sigma^2_{\varepsilon_X}}{\sigma^2_Y} + \beta^2_X )^2} \frac{1}{\sigma^2_Y} &= \frac{\frac{-\beta_X}{\sigma^2_Y}}{(\frac{\sigma^2_{\varepsilon_X}}{\sigma^2_Y} + \beta^2_X )^2}, \\
\frac{\partial}{\partial (\sigma^2_Y)}\beta_Y  & = \frac{-\beta_X}{(\frac{\sigma^2_{\varepsilon_X}}{\sigma^2_Y} + \beta^2_X )^2} \frac{-\sigma^2_{\varepsilon_X}}{(\sigma^2_Y)^2} &=\frac{\frac{\beta_X \sigma^2_{\varepsilon_X}}{\sigma^4_Y}}{(\frac{\sigma^2_{\varepsilon_X}}{\sigma^2_Y} + \beta^2_X )^2}.
\end{array}
\end{equation}

Hence, using the Delta Method~\cite[p. 61]{TheoPointEst}, the variance estimation for $\hat \beta_X$ is
\begin{equation}\label{eq:Var_BetaX}
\begin{array}{cl}
(s.e.\{\hat \beta_Y\})^2 &= (\frac{\partial}{\partial \beta_X}\beta_Y)^2 (s.e.\{\hat \beta_X\})^2 +  (\frac{\partial}{\partial (\sigma^2_{\varepsilon_X})}\beta_Y )^2(s.e.\{\hat \sigma^2_{\varepsilon_X}\})^2+ (\frac{\partial}{\partial (\sigma^2_Y)}\beta_Y)^2(s.e.\{\tilde \sigma^2_Y\})^2 \\
& =\frac{(\frac{\sigma^2_{\varepsilon_X}}{\sigma^2_Y}-\beta^2_X)^2 (s.e.\{\hat \beta_X\})^2 + (\frac{\beta^2_X}{\sigma^4_Y})(s.e.\{\hat \sigma^2_{\varepsilon_X}\})^2+ (\frac{\beta^2_X \sigma^4_{\varepsilon_X}}{\sigma^8_Y})(s.e.\{\tilde \sigma^2_Y\})^2 }{(\frac{\sigma^2_{\varepsilon_X}}{\sigma^2_Y} + \beta^2_X )^4}.
\end{array}
\end{equation}

For the quantities in the formula~\eqref{eq:Var_BetaX}, $s.e.\{\hat \beta_X\}$ can be gotten from outputs of standard linear regression fit packages, and
$$
(s.e.\{\tilde \sigma^2_Y\})^2 = \frac{2 \sigma^4_Y}{n_F-1}, \qquad (s.e.\{\hat \sigma^2_{\varepsilon_X}\})^2 = \frac{2 \sigma^4_{\varepsilon_X}}{n_s-2},
$$
from the variance formula of the chi-square distribution.

Plug-in the point estimators for each quantity into equation~\eqref{eq:Var_BetaX}, we estimate the standard error  for $\hat \beta_Y$ as
\begin{equation}\label{eq:SE_BetaY}
s.e.\{\hat \beta_Y\} = \sqrt{\frac{(\frac{\hat \sigma^2_{\varepsilon_X}}{\tilde \sigma^2_Y}-\hat \beta^2_X)^2 (s.e.\{\hat \beta_X\})^2 + (\frac{2 \hat \beta^2_X \hat \sigma^4_{\varepsilon_X}}{\tilde \sigma^4_Y})(\frac{1}{n_s-2} + \frac{1}{n_F-1})} {(\frac{\hat \sigma^2_{\varepsilon_X}}{\tilde \sigma^2_Y} + \hat \beta^2_X )^4}}.
\end{equation}

\subsection{A.2. Mathematical derivation of power formulas}

For the standard simple linear regression on a data set of size $n$, the power of an $\alpha$ level t-test (testing the zero slope null hypothesis) is given
\begin{equation}\label{eq:powerFull}
power = P( NF_{ncp=n f^2, df_1=1, df_2=n-2} > FQ_{\alpha, df_1=1, df_2=n-2}),
\end{equation}
where $NF_{ncp=n f^2, df_1=1, df_2=n-2}$ denotes a random variable following a non-central F-distribution with noncentral parameter $n f^2$, degrees of freedoms of $1$ and $n-2$, and $FQ_{\alpha, df_1=1, df_2=n-2}$ denotes the $\alpha$ upper quantile of a central F-distribution with degrees of freedoms of $1$ and $n-2$. Here $f$ is the Cohen's effect size defined as in equation~\eqref{eq:f^2}
$$
f^2 = \frac{R^2}{1-R^2} = \frac{\rho^2}{1-\rho^2},
$$
where $R^2$ is the proportion of variation in the data explained by the regression equation.

We are doing the reverse-regression hypothesis test for \eqref{eq:H0.XonY}. If we have observations of both $X$ and $Y$ on the full data set, then the power is given by formula \eqref{eq:powerFull} with $n=n_F$. Now we derive the power formula when the test is conducted on the selectively sampled subset. Besides the changes in sample size from $n_F$ to $n_S$, the effect size also changes from the full data set to the selectively sampled subset. The $R^2$ in \eqref{eq:f^2} would be larger in the subset due to selectively sampling the extreme values. To derive the change in effect size, we reexpress \eqref{eq:f^2} as
$$
f^2 = \frac{R^2}{1-R^2} = \frac{1}{\frac{1}{R^2}-1} = \frac{1}{\frac{1}{\rho^2}-1} = \frac{\beta^2_X {\sigma^2_Y}}{\sigma^2_{\varepsilon_X}},
$$
where the last equality comes from equation~\eqref{eq:rho2}. Notice that both quantities $\beta^2_X$ and $\sigma^2_{\varepsilon_X}$ remain invariant on the full data set and the subset since the reverse-regression model \eqref{eq:LR.model.XonY} holds on both data sets. $\sigma^2_Y$ differs in the two data sets.
Since only the extreme $\gamma$ proportion of $Y$ is selected, the variance of selected $Y_S$ is bigger on the selectively sampled subset than the variance of $Y$ on the full data set.
To calculate $\sigma^2_{Y_S}$, we note that the selected $Y_S$ do not follow the $N(\mu_Y, \sigma^2_Y)$ distribution anymore.  Rather it follows the normal distribution truncated at the upper and lower $(\gamma/2)$-tails. Let $z_{\gamma/2}$ denotes the upper $(\gamma/2)$-quantile of the standard normal distribution $N(0,1)$. Let $\phi(x)=\frac{1}{\sqrt{2\pi}}e^{x^2/2}$ denotes the density function of the standard normal distribution $N(0,1)$. Then the variance of selected $Y_S$ is
\begin{equation}\label{eq:VarYS}
\sigma^2_{Y_S} = \sigma^2_{Y_F} 2 \int_{z_{\gamma/2}}^{\infty} x^2 \frac{1}{\gamma} \phi(x) dx.
\end{equation}
Therefore, compared to the effect size on the full data set, the effect size $f^2$ on the subset increases by a factor of $2 \int_{z_{\gamma/2}}^{\infty} x^2 \frac{1}{\gamma} \phi(x) dx$. Since the sample size $n_S$ of the subset is $\gamma$ proportion of the full data set size $n_F$ so that $n_S/\gamma=n_F$, the power of t-test on the subset becomes
\begin{equation}\label{eq:powerSel}
\begin{array}{cl}
power & = P( NF_{ncp=n_S f^2 2 \int_{z_{\gamma/2}}^{\infty} x^2 \frac{1}{\gamma} \phi(x) dx, df_1=1, df_2=n_S-2} > FQ_{\alpha, df_1=1, df_2=n_S-2}) \\
& = P( NF_{ncp=n_F f^2 2 \int_{z_{\gamma/2}}^{\infty} x^2 \phi(x) dx, df_1=1, df_2=\gamma n_F-2} > FQ_{\alpha, df_1=1, df_2=\gamma n_F-2}).
\end{array}
\end{equation}
%Using this formula, we can plot the power versus the selected proportion $\gamma$.

\end{document}